\documentclass[namedreferences]{solarphysics}

\usepackage[hyperref,optionalrh,showbiblabels]{spr-sola-addons} 
\usepackage{graphicx}        
\usepackage{color}           
\usepackage{breakurl}        
\usepackage{rotating}
\usepackage{lscape}
\usepackage{enumerate}
\usepackage{tikz}
\usetikzlibrary{decorations.pathreplacing}
\usetikzlibrary{shapes.geometric, arrows}
\tikzstyle{flowbox} = [rectangle, rounded corners, minimum width=7cm, minimum height=1cm,  text width=7cm, text centered, draw=blue]
\tikzstyle{arrow} = [thick,->,>=stealth]
\usepackage{float}

\newcommand{\etal}{{et al.}}

\usepackage{multirow}

\hypersetup{colorlinks = true, linkcolor = red, anchorcolor = red, citecolor = blue, filecolor = red, pagecolor = red, urlcolor = red}
\ifx\urlurl    \undefined
	\def\urlurl#1{\href{https://#1}{\textsf{#1}}}\fi

\chardef\us=`\_
\newcommand{\edcom}[1]{#1}

\begin{document}

\begin{article}
\begin{opening}

\title{\Large A Simple Radial Gradient Filter for Batch-Processing of Coronagraph Images}

\author[addressref={aff1,aff1a,aff1b}]{\inits{R.}\fnm{Ritesh}~\lnm{Patel}\orcid{0000-0001-8504-2725}}
\author[addressref={aff1,aff1c}]{\inits{S.}\fnm{Satabdwa}~\lnm{Majumdar}\orcid{0000-0002-6553-3807}}
\author[addressref={aff1a},email={vaibhav.pant@aries.res.in}]{\inits{V.}\fnm{Vaibhav}~\lnm{Pant}\orcid{0000-0002-6954-2276}}
\author[addressref={aff1,aff1a,aff3},corref,email={dipu@iiap.res.in}]{\inits{D.}\fnm{Dipankar}~\lnm{Banerjee}\orcid{0000-0003-4653-6823}}

\address[id=aff1]{Indian Institute of Astrophysics, Koramangala, Bangalore, 560034, India}
\address[id=aff1a]{Aryabhatta Research Institute of Observational Sciences, Nainital 263001, India}
\address[id=aff1b]{University of Calcutta, 87/1, College Street, Kolkata, 700073, India}
\address[id=aff1c]{Pondicherry University, Chinna Kalapet, Kalapet, Puducherry 605014, India}
\address[id=aff3]{Center of Excellence in Space Sciences, IISER Kolkata - 741246, India}

\runningauthor{{R. Patel \etal}}
\runningtitle{\edcom{Simple Radial Gradient Filter}}
\begin{abstract}
{Images of the extended solar corona, as observed by white-light coronagraphs as observed by different white-light coronagraphs include the  K- and F-corona and suffer from a radial variation in intensity.} These images require separation of the two coronal components with some additional image-processing to reduce the intensity gradient and analyse the structures and processes occurring at different heights in the solar corona within the full field of view. Over the past few decades, coronagraphs have been producing enormous \edcom{amounts} of data, which will be continued with the launch of new telescopes. To process these bulk coronagraph images with steep radial-intensity gradients, we have developed an algorithm: {\it Simple Radial Gradient Filter} (SiRGraF). This algorithm is based on subtracting a minimum background (F-corona) created using long duration images and then dividing the resultant by a uniform intensity gradient image { to enhance the K-corona}. { We demonstrate the utility of this algorithm to bring out the short time-scale transient structures of the corona.} { SiRGraF could be used to reveal and analyse such structures. It is not suitable for quantitative estimations based on intensity.}
We have successfully tested the algorithm on images of the {\it Large  Angle  Spectroscopic COronagraph} (LASCO)-C2 \edcom{onboard} the {\it Solar and Heliospheric Observatory} (SOHO), and COR-2A \edcom{onboard} the {\it Solar TErrestrial RElations Observatory} (STEREO) with good signal to noise ratio (SNR) along with low-SNR images of STEREO/COR-1A and the {\it KCoronagraph} (KCor). We also compared the performance of SiRGraF with an existing widely used algorithm: {\it Normalising Radial Gradient Filter} (NRGF). We found that when hundreds of images have to be processed, SiRGraF works faster than NRGF, providing similar brightness and contrast in the images and {separating the transient features}. Moreover, SiRGraF works better on low-SNR images of COR-1A than NRGF, providing better identification of dynamic coronal structures throughout the field of view. We discuss the advantages and limitations of the algorithm. The application of SiRGraF to COR-1 images could be extended for an automated coronal mass ejection (CME) detection algorithm in the future, which will help in our study of the CMEs' characteristics in the inner corona.

\end{abstract}
\keywords{Corona; Corona, Structures; Instrumentation and Data Management}
\end{opening}

%
\section{Introduction} 
    \label{sec: intro} 

{ The observed white-light corona \edcom{contains} of K- and F-components formed due to scattered photospheric light by the free electrons and dust particles, respectively, with the later dominating beyond $\approx$\,3\,R$_\odot$ \citep{Hulst50, Morgan07}}. The shape of the white-light corona is governed by the density structures where the K-corona is associated with the magnetic-field configuration \citep{Woo2005SoPh}. { It has been found that the F-corona remains virtually constant throughout the solar cycle at heights more than $\approx$\,2.6\,R$_\odot$ \citep{Morgan07}. Large-scale transients, \edcom{coronal mass ejections} (CMEs), observed in white-light images are electron-density structures \citep{Howard11} that require a careful separation from the background \citep{Byrne12}. A general method to separate the F-corona is to subtract the minimum intensity obtained over a long period of time \citep{DeForest14, DeForest2018ApJ} or \edcom{use} a monthly minimum of daily median images \citep{Morrill2006SoPh}. \citet{Morgan2010ApJ} separated the transient component by estimating a background based on polynomial fitting in each radial direction. It was further developed using iterative methods for segregating the quiescent and dynamic corona while automatically detecting CMEs \citep{Morgan12}.} { Later \citet{Morgan2015} introduced a methodology based on radiometric calibration of LASCO-C2 images to subtract the background with suppressing the noise while separating the K-corona and the dynamic structures.}

One of the challenges in coronal observations,{ apart from separation of the two coronal components}, is that due to the decreasing electron and particle density with height there is a steep gradient in the observed intensity \citep{Baumbach37, Hulst50}. { The radial variation of the F- and K-corona has been quantified by \citet{Morgan07} over different phases of the solar cycle.} The analysis using coronagraph images becomes challenging unless some image-processing technique is used to reduce the brightness variation. Various methods, in the form of radial gradient filter (RGF) have been developed to overcome this difficulty. In this regard, RGF-based photography was \edcom{made} using a mechanical rotating sector to image the corona \citep{Owaki1967} or a radial density filter \edcom{was} placed in the optical path of \edcom{an} imaging instrument to photograph solar eclipses \citep{Newkirk1968SoPh}. Solar eclipse photographs were made as composite images, combining multiple exposure frames for the process. To enhance the high dynamic range of brightness in the corona during such event, different image-processing techniques were developed \citep{Koutchmy1992A&A, Espenak2000, Druckm2006}. These techniques enhance the structures with high spatial frequencies leading to the visibility of sharp features in the images. To identify the CMEs in the midst of such dynamic coronal intensity, \citet{Byrne2009A&A} introduced a method based on multi-scale filtering, thereby enhancing the visibility of CMEs.

\citet{NRGF2006SoPh} introduced \edcom{the} normalising-radial-gradient filter (NRGF), which subtracts the mean intensity followed by division \edcom{by the} standard deviation at each height to reduce the radial-intensity variation. This method has been extensively used to enhance the coronal structures with applications to eclipse images \citep{Pasachoff_2007, Habbal2010ApJ, Habbal2011ApJ, Boe2018Freezin}, automated detection of CMEs using the {\it Coronal Image Processing} (CORIMP) technique \citep{Byrne12, Morgan12}, Automated CME Triangulation \citep{ACT2017A&A}, identification of plasmoids in the corona \citep{Lee2020ApJ, patel2020A&A}, and identification of Alfv\'en waves in the solar atmosphere \citep{He2009A&A}.
An improvement on this algorithm was developed \edcom{known as} the {\it Fourier normalising-radial-graded filter} (FNRGF), based on finite Fourier series that takes a local average and standard deviation, \edcom{and this} was used to enhance the fine coronal structures in the low contrast regions \citep{FNRGF2011}. Recently, a new method called the {\it radial local multi-scale filter} (RLMF) was developed using multi-scale filtering on radial \edcom{vectors} extracted from coronagraph images followed by intensity normalisation leading to enhancement of coronal structures \citep{RLMF2020101383}. For the detection of inbound waves in the solar corona, \citet{DeForest14} normalised the radial intensity of the COR-2 images by subtracting the average intensity across a column and time from each row, followed by dividing each row by its standard deviation across column and time. 

{Methods based on wavelets to enhance coronal features in extreme ultraviolet (EUV) images were developed by \citet{Stenborg2003A&A} and \citet{Stenborg2008ApJ}.} To enhance the off-disk coronal structures in EUV images of the {\it Atmospheric Imaging Assembly} \citep[AIA:][]{AIA} by improving the intensity variation radially, {IDL routines \textsf{aia\_rfilter.pro} and \textsf{aia\_rfilter\_jp2gen.pro} were developed ({see \urlurl{aia.cfa.harvard.edu/software.shtml}}). These add the off-limb component of many AIA images thereby increasing the signal-to-noise ratio (SNR). The off-disk corona is then further divided into
concentric rings, which are scaled as a function of its radial distance, average intensity, and the intensity with respect to rings in the neighborhood \citep{Masson2014ApJ}. EUV coronal structures were also enhanced while reducing the dominance of noise based on a fuzzy algorithm \citep{NAFE2013ApJS} and using Gaussian filtering for bringing out features at different scales \citep{Morgan2014SoPh}.}
In another algorithm, {\it CMEs Identification in Inner Solar Corona} (CIISCO), a radial filter was applied on EUV images by dividing \edcom{each} individual image by a background \edcom{uniform in azimuth direction} generated using the radial intensity profile of the polar regions of the Sun \citep{ciisco2020}.

Each of the above-mentioned algorithms has its own advantages and limitations. Even though all of these succeed in enhancing the coronal structures and reducing the intensity variation, the processing time becomes large when the number of images to be analysed becomes large. Bulk image-processing is required for \edcom{such} cases, some of \edcom{that} include long-term study of CMEs, and long-term study of solar corona \citep{Lamy2020c2, Lamy2020c3}. { Therefore in this article we present an algorithm, {\it Simple Radial Gradient Filter} (SiRGraF), which works faster when hundreds of images have to be processed to reveal the dynamic coronal features}. The article is arranged as follows: the algorithm is introduced in Section \ref{sec: algo}, \edcom{and} the results of the application of SiRGraF on coronagraph images and comparison with NRGF in Section \ref{sec: results}. We then summarise our work and discuss the analysis in Section \ref{sec: summary}.
 
\section{Algorithm} 
    \label{sec: algo}
The observations of the white-light corona in coronagraph images \edcom{exhibit} a steep \edcom{outward} \citep{Hulst50, NRGF2006SoPh, Morgan07}. To analyse coronagraph images and study the associated physical process specially transient activities in detail, it is important to identify the coronal features throughout the FOV. { In this work we have used the Level-1 images of the white-light coronagraphs including {\it Large Angle Spectrometric Coronagraph} \citep[LASCO:][]{Brueckner95} onboard the {\it Solar and Heliospheric Observatory} (SOHO), {\it Sun Earth Connection Coronal and Heliospheric Investigation} \citep[SECCHI:][]{Howard2008SSRv} of the {\it Solar Terrestrial Relations Observatory} (STEREO), and the {\it KCoronagraph} \citep[KCor:][]{deWijn} of the Mauna Loa Solar Observatory (MLSO). The Level-1 images are the total brightness images calibrated to the mean solar disk brightness, flat and dark corrected with alignment and \edcom{with} corrections done for solar north up.}
(
{ We demonstrate in the following steps the methodology involved in the SiRGraF, using Level-1 STEREO/COR-1A data of \edcom{01 August 2010} taken at a cadence of \edcom{five} minutes:}

\begin{figure}[!ht]   
   \centerline{\hspace*{0.05\textwidth}
               \includegraphics[width=0.58\textwidth,clip=]{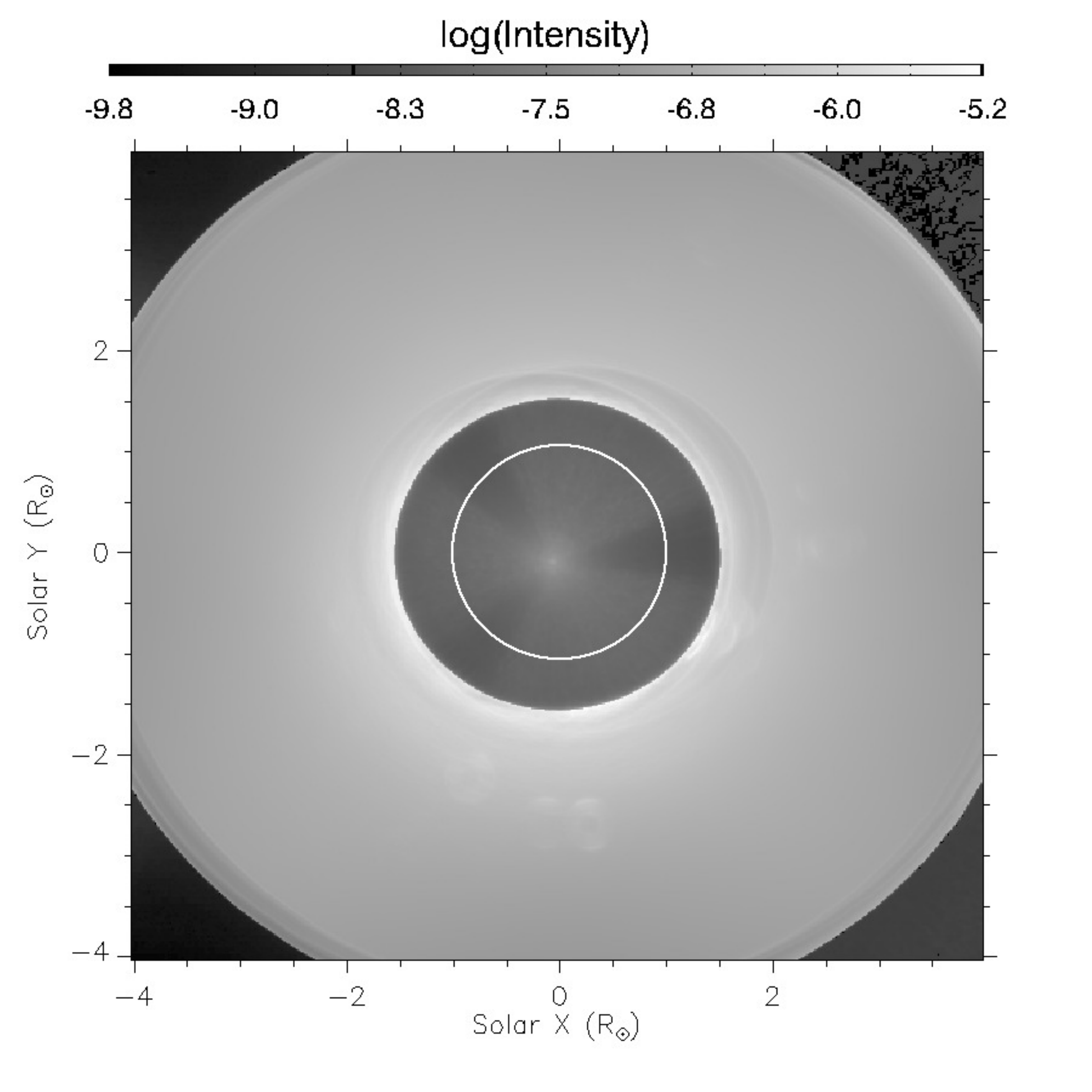}
               \hspace*{0.002\textwidth}
               \includegraphics[width=0.58\textwidth,clip=]{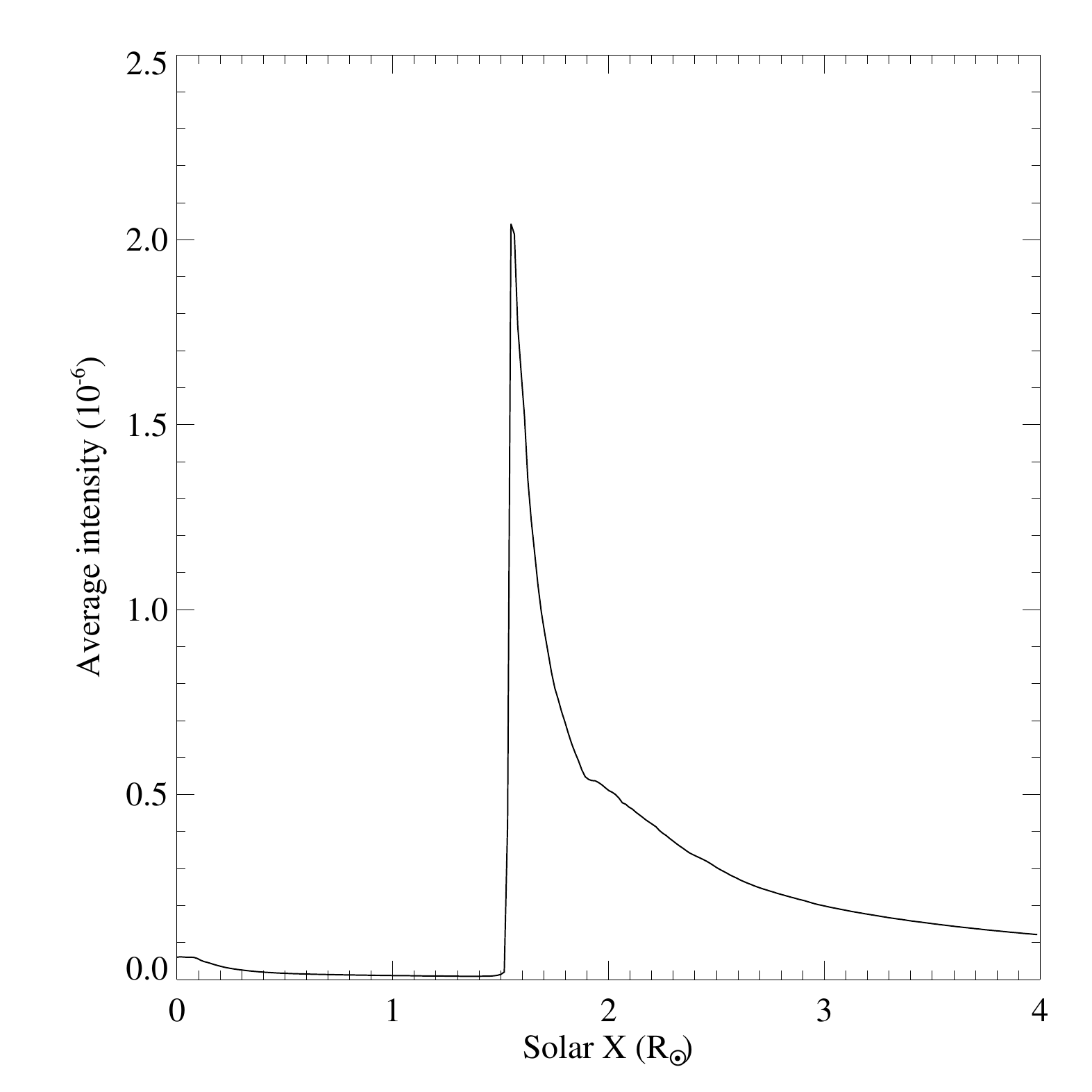}
              }
      \vspace{-0.01\textwidth}  
     \centerline{    
      \hspace{0.2\textwidth}  \color{black}{(a)}
      \hspace{0.58\textwidth}  \color{black}{(b)}
         \hfill}
     \vspace{0.005\textwidth}     
          
 \centerline{\hspace*{0.05\textwidth}
               \includegraphics[width=0.58\textwidth,clip=]{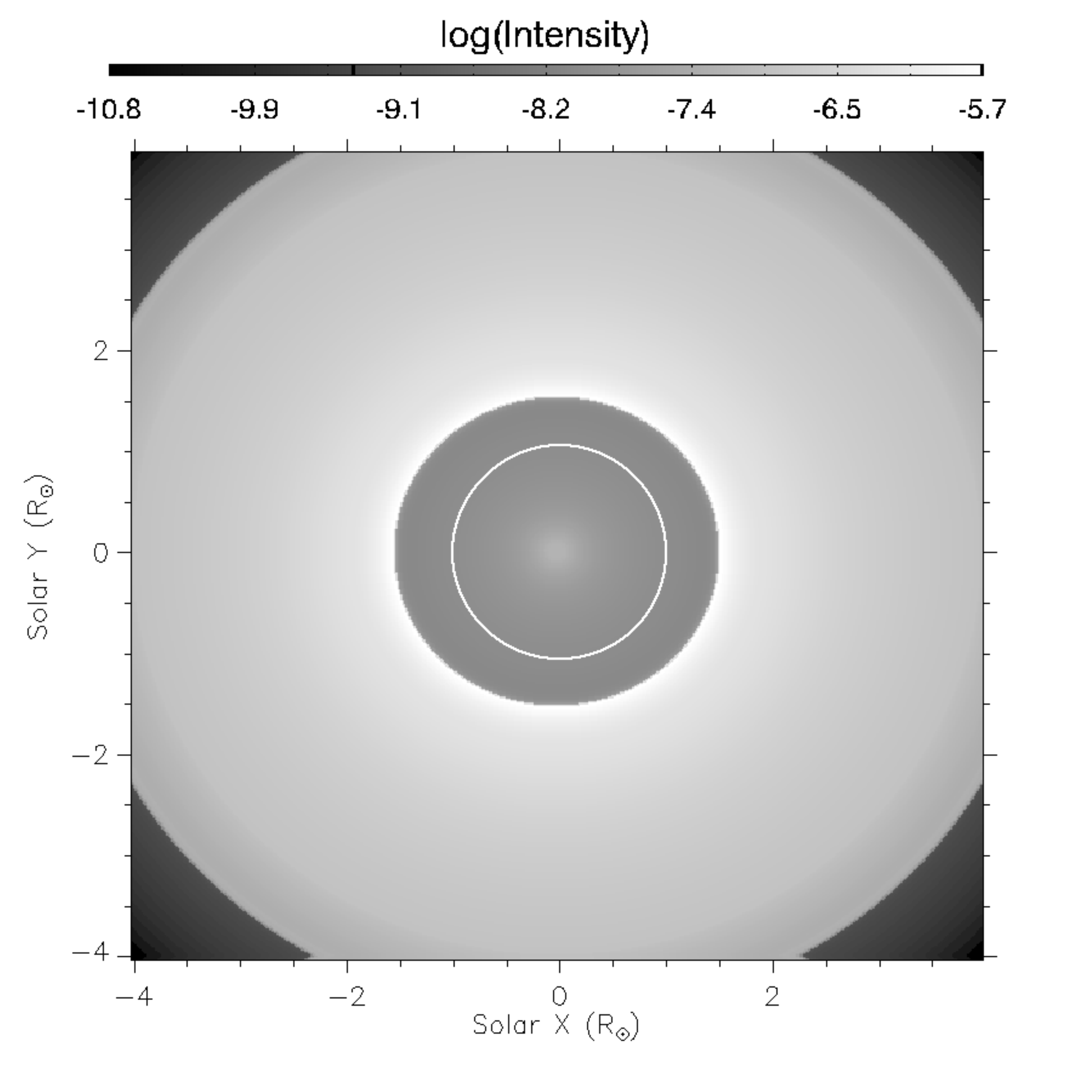}
               \hspace*{0.002\textwidth}
               \includegraphics[width=0.58\textwidth,clip=]{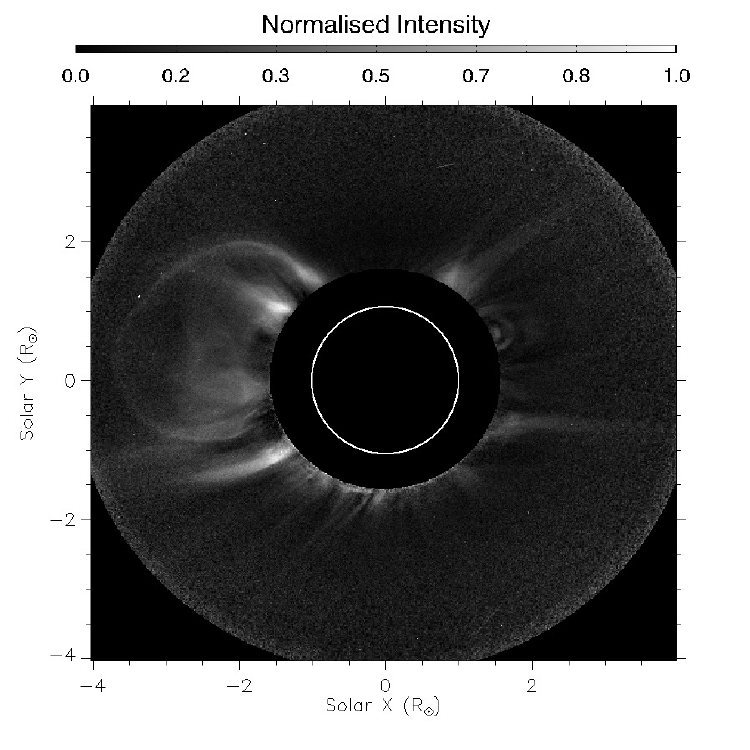}
              }
     \vspace{-0.01\textwidth}  
     \centerline{   
      \hspace{0.2\textwidth}  \color{black}{(c)}
      \hspace{0.58\textwidth}  \color{black}{(d)}
         \hfill}
     \vspace{0.01\textwidth} 
\caption{An outline of the SiRGraF algorithm on application to images of STEREO/COR-1A observed on \edcom{01 August 2010}; ({\bf a}) Minimum background image created from Level-1 images of whole day, ({\bf b}) Azimuth average radial-intensity plot generated from the minimum background image, ({\bf c}) A circularly symmetric uniform background generated from the radial-intensity array of ({\bf b}), ({\bf d}) Final image at 08:27\,UT after subtracting the minimum background and dividing the resultant image by uniform background. { The intensity in ({\bf a}) and ({\bf c}) are displayed on a log scale while normalised intensity is displayed in ({\bf d}).}}
   \label{fig:algo_op}
\end{figure}

\begin{enumerate}[i)]
    \item Level-1 images [$I$] of a single day are taken to produce a single image, called minimum background [$I_{\rm m}$], as shown in Figure \ref{fig:algo_op}a. This is generated such that each pixel in $I_{\rm m}$ corresponds to the minimum intensity { (with intensity greater than zero to avoid the outliers)} of all the images. This image consists of intensity from F-corona, less variable K-corona, and instrumental scattered light \citep{Morgan07}. { The minimum background is taken instead of the 1 percentile minimum to make the processing fast and efficient.}
    \item { This minimum background is used to generate a radial-intensity profile such that the intensity at a certain height is an average of all of the intensities in the azimuthal direction at that height (Figure \ref{fig:algo_op}b). This smooths out the variation intensity in the azimuth direction. As one-day minimum background is used here, the presence or absence of long-lived streamers may contribute at those \edcom{distances above the limb}, which gets averaged out while creating the radial 1D-profile.}
    \item The radial profile is then incorporated to produce a uniform background image [$I_{\rm u}$] with a circularly symmetric intensity gradient. Figure \ref{fig:algo_op}c shows a uniform background image made using the radial profile shown in Figure \ref{fig:algo_op}b. As the instrument background and scatter remains nearly uniform from Equator to polar regions  \citep{Morgan07, Patel2018}, this will serve to normalise such contributors of the radial gradient along with inherent coronal-intensity variation.
    \item After obtaining the background images, the Level-1 images are thereby filtered for the radial-intensity gradient using the following relation:
    
    \begin{equation} \label{eq: sirgraf}
        I' = \frac{I-I_\mathrm{m}}{I_\mathrm{u}}.
    \end{equation}
\end{enumerate}

Equation \ref{eq: sirgraf} represents the core of the SiRGraF. The numerator shows the removal of the static and quasi-static components of the corona along with instrumental background bringing out the K-corona. { This minimum background subtraction is in contrast with the pipeline of the LASCO and STEREO coronagraphs where monthly minimum \edcom{background} of daily median images are used as background. One should note that the daily median image has a contribution from K-corona more than the daily minimum image \citep{Thompson2010SoPh}. For the objective of separating the dynamics and transients (CMEs), it is required to capture the maximum part of the K-corona signal.}
When the K-corona obtained after removing the background is divided by a uniform background, the radial variation of intensity is reduced, thereby allowing visualisation of coronal structures to greater heights uniformly in the azimuth direction. The filtered images with normalised intensities will thus visibly exhibit coronal structures as shown in Figure \ref{fig:algo_op}d. 
{ The use of single-day minimum background brings out the more dynamic components of the K-corona, which in this case are the CMEs and dynamically changing streamers as seen in Figure \ref{fig:algo_op}d. The streamers on this day were either displaced by the CMEs or short-lived making them dynamic in nature. The use of a extended-period minimum background could bring out long-lived structures of the corona, which could not be extracted in the example presented here.}

\section{Results}
    \label{sec: results}

\subsection{Application to Coronagraph Images}
\label{sec:coronagraphs}

\begin{figure}[!htbp]   
   \centerline{\hspace*{0.05\textwidth}
               \includegraphics[width=1.5\textwidth,clip=]{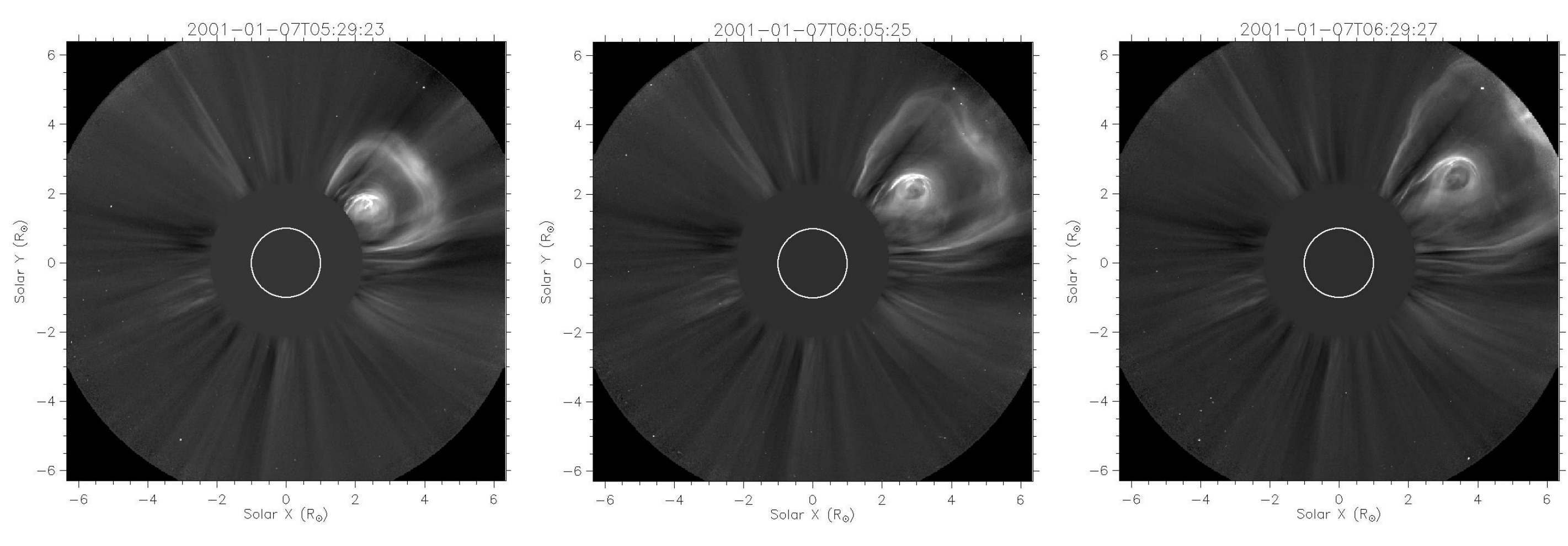}
                }
      \vspace{-0.01\textwidth} 
    \centerline{\hspace*{0.05\textwidth}
               \includegraphics[width=1.5\textwidth,clip=]{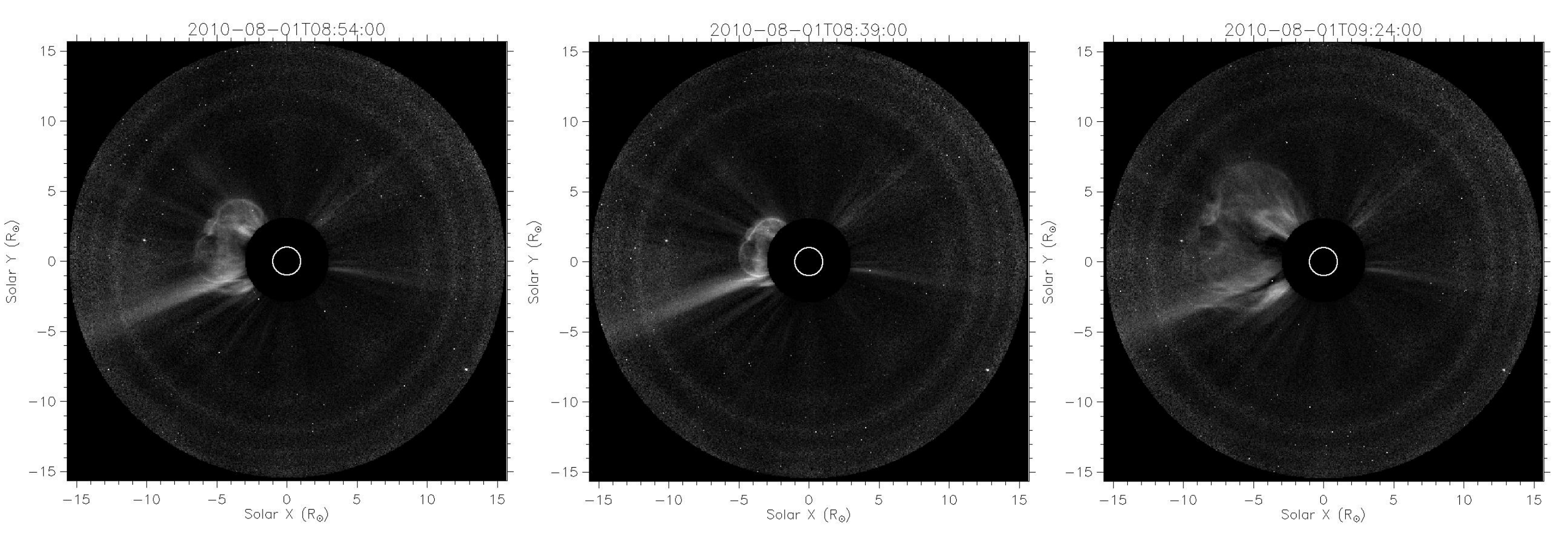}
                }
      \vspace{-0.01\textwidth} 
    \centerline{\hspace*{0.05\textwidth}
               \includegraphics[width=1.5\textwidth,clip=]{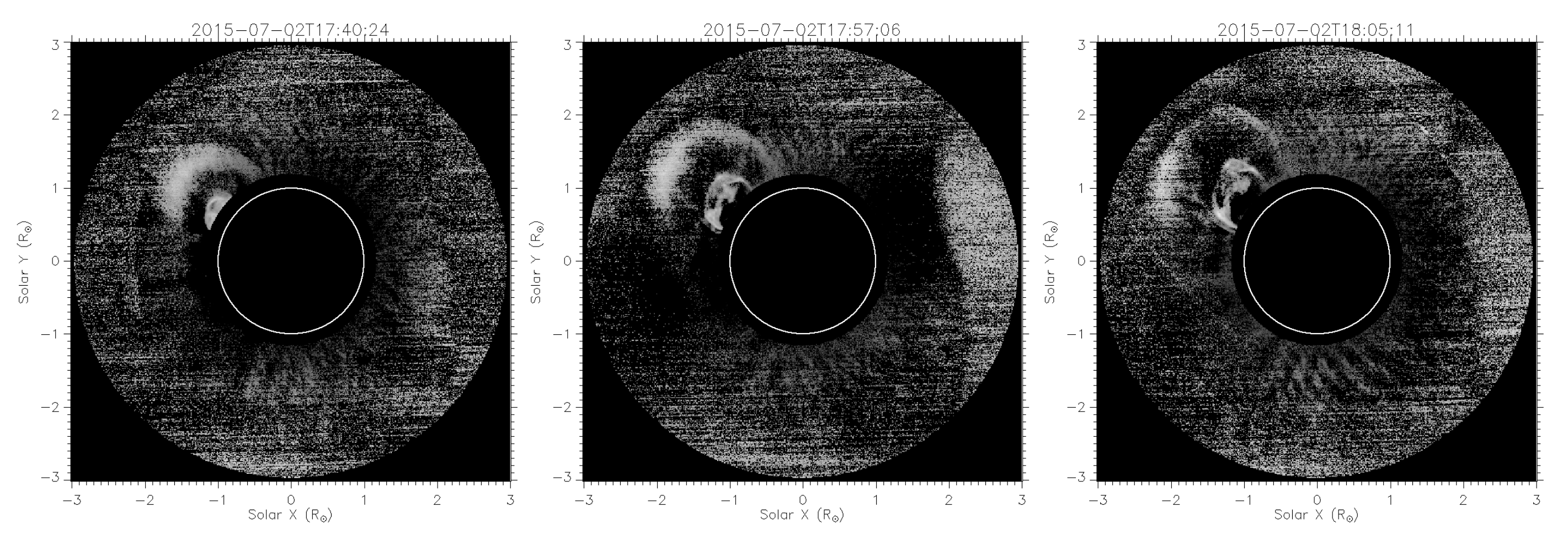}
                }
      \vspace{-0.01\textwidth}   
\caption{Application of SiRGraF on successive images of LASCO-C2 taken on \edcom{07 July 2001}, STEREO/COR-2A taken on \edcom{01 August 2010}, and KCor taken on \edcom{02 July 2015} respectively in {\it top, middle and bottom panels}. Different parts of the CME and coronal structures could be clearly seen to the { outer edge} of FOV for C2 and COR-2A images and only CME structures in KCor images. The images are displayed with normalised intensity.}
   \label{fig:c2filt}
\end{figure}

An application of SiRGraF \edcom{to} STEREO/COR-1A images is illustrated in Figure~\ref{fig:algo_op}. We also tested the algorithm on {Level-1 images of other white-light coronagraphs} of LASCO-C2, STEREO/COR-2A, and MLSO/KCor. Figure \ref{fig:c2filt} shows such an example where top panel shows radial-filtered LASCO-C2 images of \edcom{07 January 2001} using SiRGraF. { The backgrounds for the space-based coronagraphs have been prepared using the available whole-day images.} It can be seen that the different parts of the classic three-part-structured CME have been brought out clearly by SiRGraF throughout the FOV. It should be noted that an example of this CME using NRGF has been presented \edcom{by} \citet{NRGF2006SoPh} and a direct comparison could be made in the two processes. The middle panel of the same \edcom{figure} shows filtered successive images of STEREO/COR-2A of \edcom{01 August 2010,} which is the same CME example taken for COR-1A. For this instrument, CME and other coronal structures could be distinctly identified to considerably larger heights. { We could also notice faint rings near the outer edge of the FOV in COR-2A images. These rings are due to increased photon noise near the outer vignetting minimum of the instrument \citep{DeForest2018ApJ}.} These rings have intensity greater than the backgrounds we have used for filtering and hence are visible in the processed images. It should be noted that these are of instrumental origin and \edcom{it} will require advance processing to remove these artifacts \citep{DeForest2018ApJ}.

We also applied this algorithm on Level-1 images of KCor taken on \edcom{02 July 2015} with cadence of 15 \edcom{seconds}. { As KCor is ground-based coronagraph, we employed \edcom{only} the limited hours of observed data available for this day to create the required backgrounds.} In the bottom panel of Figure \ref{fig:c2filt} the output after the implementation of the algorithm could be seen in KCor images for three instances. It can be seen that inspite of the atmospheric contributions, the different parts of the CME can be identified almost to the { outer edge} of the FOV. Such enhancement is achieved without the need to add the images to improve the SNR or compromising on the cadence. KCor being a ground-based coronagraph, the minimum background suffers from the atmospheric contamination and hence could not be obtained as clearly as for space-based coronagraphs. As a result, the signatures of the streamers are lost for this case, limiting SiRGraF to mostly space-based coronagraph images and to only analysis involving erupting structures for ground-based coronagraph images. A longer period (ten days to one month) for minimum background creation could be considered for this instrument in the future for significantly bringing out the signal of streamer-like long-lived structures.

\subsection{Comparison with Extended Period Background}
\label{sec:7daymin}

\begin{figure}[!htbp]   
   \centerline{\hspace*{0.05\textwidth}
    \includegraphics[width=0.7\textwidth,clip=]{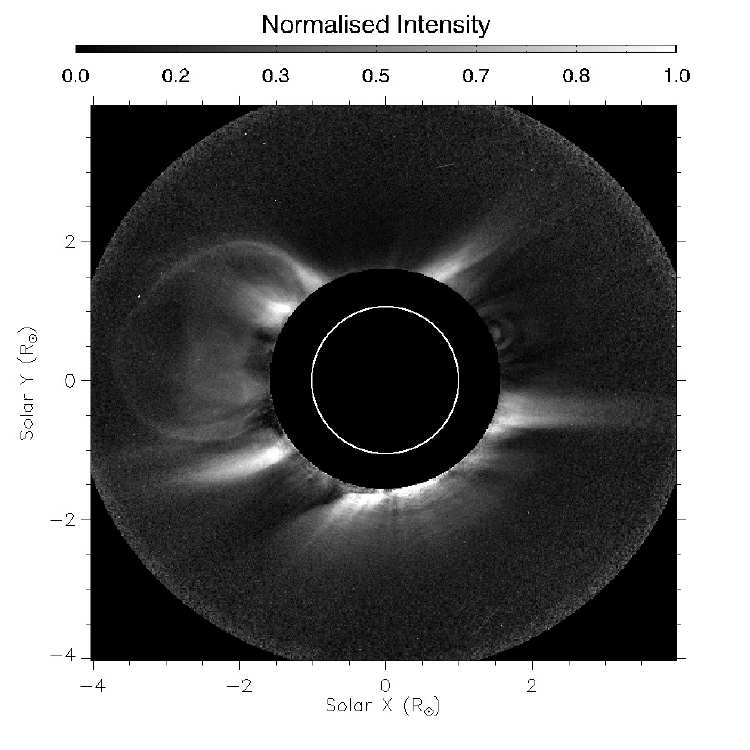}
    \includegraphics[width=0.7\textwidth,clip=]{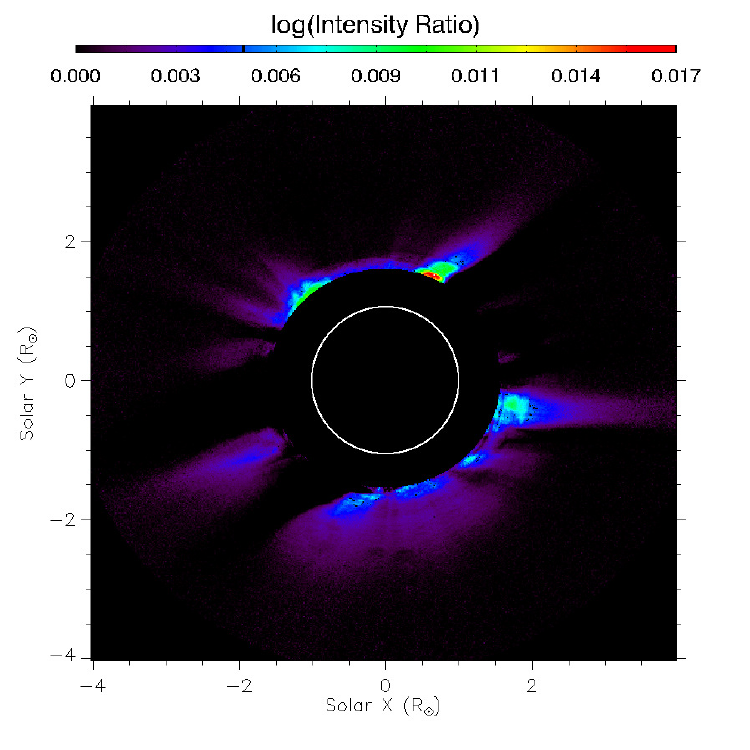}
     }
      \vspace{-0.01\textwidth}  
      \centerline{    
      \hspace{0.15\textwidth}  \color{black}{(a)}
      \hspace{0.675\textwidth}  \color{black}{(b)}
         \hfill}
\caption{({\bf a}) COR-1A images of \edcom{01 August 2010} at 08:27\,UT processed with SiRGraF using backgrounds created with seven days intensity images, ({\bf b}) Intensity ratio of minimum backgrounds using one-day and seven-day images.}
   \label{fig:7day_comp}
\end{figure}

{ We also created a seven-day background to identify the effects of long-period background while processing the coronagraph images. We used total-brightness images from \edcom{26 July 2010} to \edcom{01 August 2010} to process the COR-1A images of \edcom{01 August 2010}. The minimum background was created using this extended period of data and the uniform background was obtained based on this minimum image as mentioned in Section \ref{sec: algo}. Figure \ref{fig:7day_comp}a shows the output of the algorithm after application. When it is compared with Figure \ref{fig:algo_op}d, we can see that the transient CME is brought out clearly in both cases. In addition to CMEs, the streamers are more distinctly visible when an extended background is used in the algorithm. The streamers located near the north-west and western limb of the Sun are more pronounced, as suggested in Section \ref{sec: algo}.

We also compared the intensities of the minimum background created using one-day and seven-day images for the same dataset (Figure \ref{fig:7day_comp}b). One can notice that the background plays a small role for the majority of the regions where CME is present.
The ratio suggests an increased contribution from streamers is present for a single-day background than for the seven-day case. Hence, it is also reflected in the processed images. Even though this ratio only accounts for $\approx$\,5\,\% difference in the background determination, but its effect could be seen when long-lived structures such as streamers are considered. One needs to identify the science target while considering the period for creating the backgrounds. 
}

\subsection{Comparison with NRGF}
\label{sec:nrgfcompare}
 {The SiRGraF algorithm uses Equation \ref{eq: sirgraf} while NRGF is based on the following relation:}
\begin{equation}
    I_{\mathrm{filt}}(r, \phi) = \frac{I(r, \phi)-I(r)_{\langle\phi\rangle}}{\sigma(r)_{\langle\phi\rangle}},
    \label{eq:nrgf}
\end{equation}
{where $I_{\rm filt}(r, \phi$) is the NRGF processed image, $I(r, \phi$) is the original intensity image at height $r$ and position angle $\phi$, $I(r)_{\langle\phi\rangle}$ and $\sigma(r)_{\langle\phi\rangle}$ are the average and standard deviation of the intensities at the height {\it r} computed over all position angles. On comparing Equations \ref{eq: sirgraf} and \ref{eq:nrgf}, we see that SiRGraF performs matrix operation using the backgrounds at a \edcom{given} time whereas NRGF \edcom{performs row-wise operation} requiring the computation of mean and standard deviation at each height at a \edcom{given} time. Both \edcom{of} these methods subtract an average intensity, which is then normalised to flatten the radial-intensity variation.} We used 339 Level-1 total brightness images of STEREO/COR-1A of size 512$\times$512 pixels and processed with SiRGraF and NRGF on an i7 \edcom{computer} with a base clock speed of 1.8\,GHz and 8\,GB of RAM. { The application of SiRGraF is based on the steps mentioned in Section \ref{sec: algo} where the backgrounds are created using all 339 images. We used the IDL routine \textsf{nrgf.pro} to process the images with NRGF.} 



\begin{figure}[!htbp]   
   \centerline{\hspace*{0.05\textwidth}
               \includegraphics[width=0.58\textwidth,clip=]{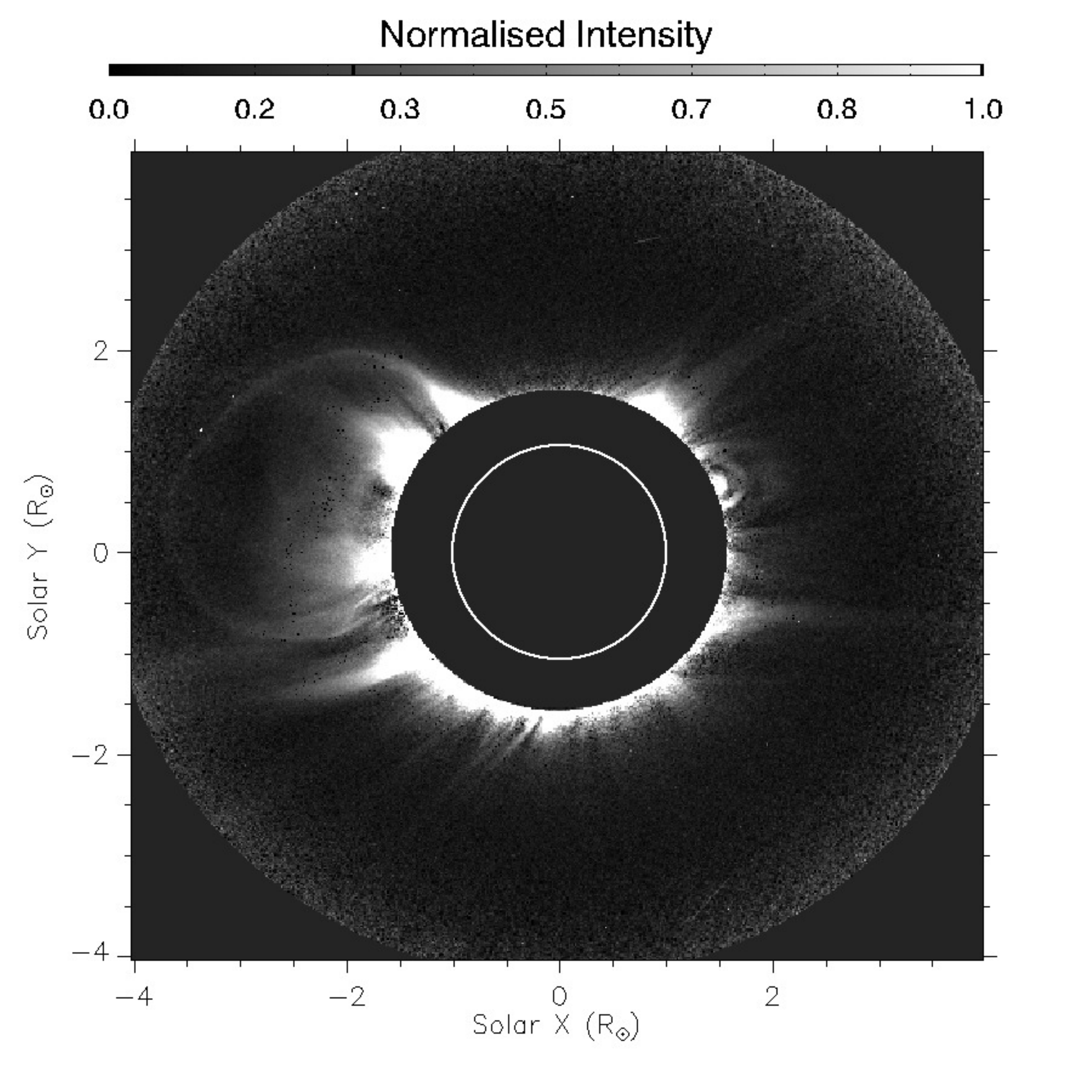}
              }
      \vspace{-0.01\textwidth}  
%
\caption{Output of application of NRGF on COR-1A image of 01 August 2010 08:27\,UT.  An animation is available in the \edcom{Electronic Supplementary Material}.}
   \label{fig:nrgf}
\end{figure}

{ We found that SiRGraF takes a comparatively longer time { ($\approx$\,2.5 seconds)} to process a single image whereas NRGF takes only a fraction of a second. This is because the two backgrounds that need to be generated for SiRGRaF take the majority of the time in the process. At the same time, it should also be kept in mind that SiRGraF produces images where the coronal features are visually identified with relatively better clarity and ease. When the whole batch of 339 images is processed by the two filters, SiRGraF takes $\approx$\,three seconds to complete the whole process similar to NRGF when the inputs to Equation \ref{eq:nrgf} are kept constant over the batch of images. This defines the efficiency of this algorithm when hundreds of images have to be processed. 
We used the Michelson contrast ratio \citep{michelson1927studies} to compare the intensity contrast in processed images. This parameter is defined as,
    \begin{equation}
        V = \frac{I_{\mathrm{max}}-I_{\mathrm{min}}}{I_{\mathrm{max}}+I_{\mathrm{min}}},
    \end{equation}
where $I_{\rm max}$ and $I_{\rm min}$ are the maximum and minimum intensities in the image. We measured the contrast in individual images and then averaged for the whole batch. It turns out that the mean contrast produced by the two methods is close to unity and hence \edcom{they} are very similar in nature.}

After going through the quantitative assessments of the two methods, we looked for the visual differences in the outputs of the two. An animation is also available with Figure \ref{fig:nrgf} providing a side-by-side comparison of SiRGraF with NRGF \edcom{applied} to COR-1A images.
{ On comparing Figure \ref{fig:nrgf} with Figure \ref{fig:algo_op}d visually along with the animation, we found that both SiRGraf and NRGF reveal the coronal structures upto the { outer edge} of FOV in the images. We noticed that SiRGraF-processed COR-1 images appear to be uniformly illuminated as compared to the NRGF ones. In NRGF-processed images the features in the inner FOV appears brighter than those at the { outer edges}. Due to this the structures at the inner FOV in NRGF processed images are not distinctly visible when compared with SiRGraF processed ones.}

\subsection{Solar-Cycle Variations}   
\label{sec:solcycle}

\begin{figure}[!htbp]   
   \centerline{\hspace*{0.05\textwidth}
               \includegraphics[width=1.2\textwidth,clip=]{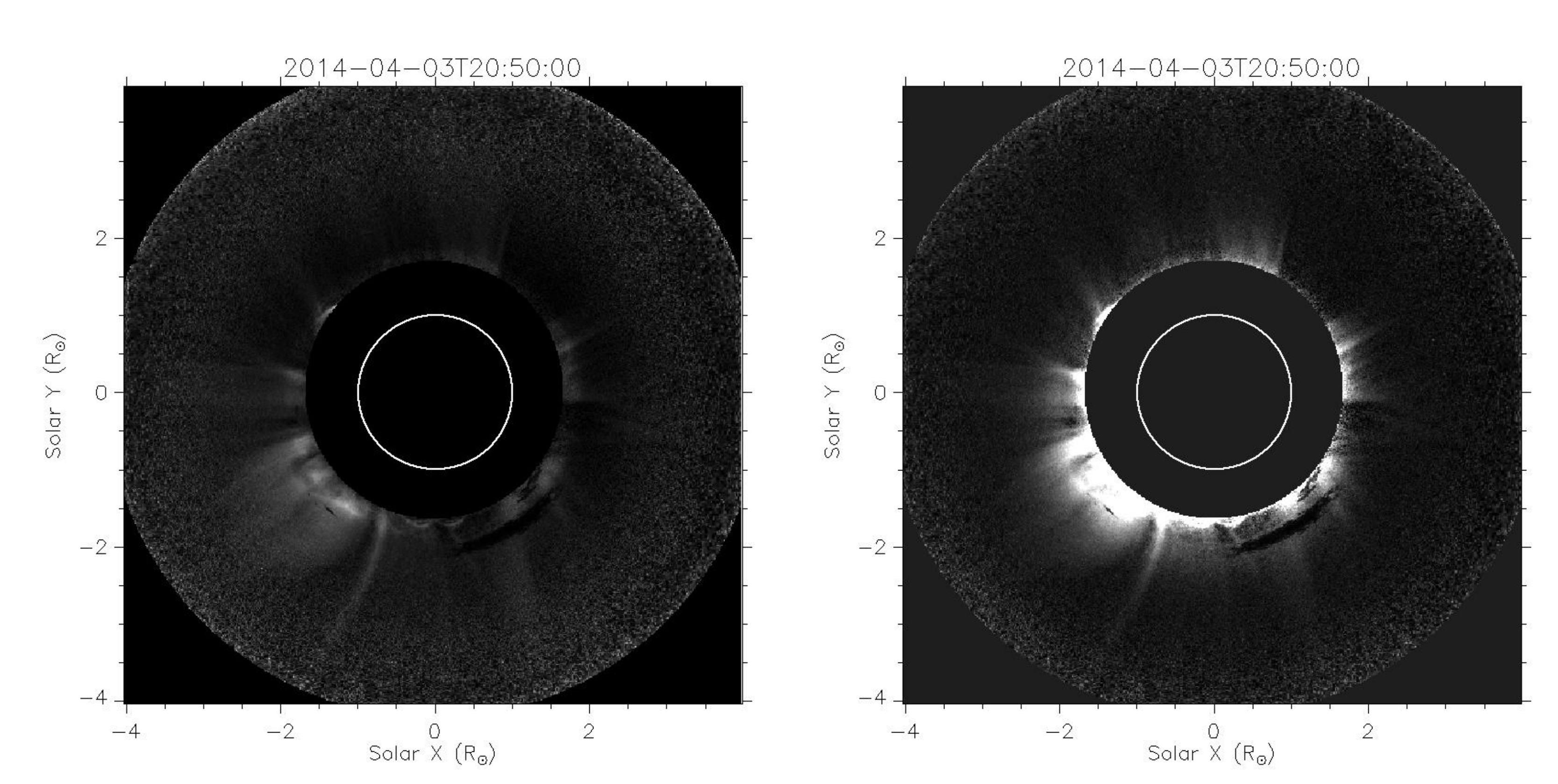}
                }
      \vspace{-0.01\textwidth} 
    \centerline{\hspace*{0.05\textwidth}
               \includegraphics[width=1.2\textwidth,clip=]{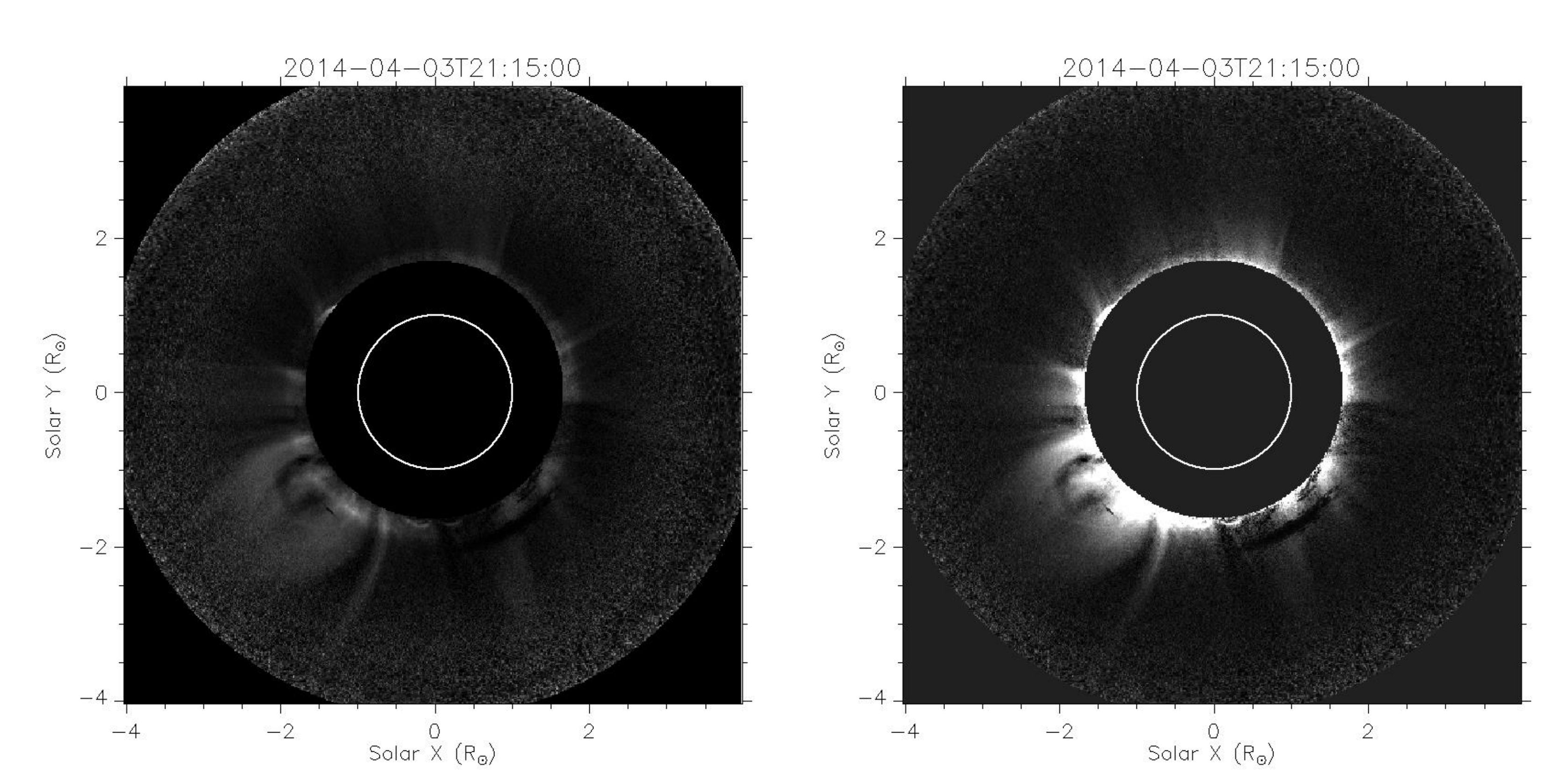}
                }
      \vspace{-0.01\textwidth} 
    \centerline{\hspace*{0.05\textwidth}
               \includegraphics[width=1.2\textwidth,clip=]{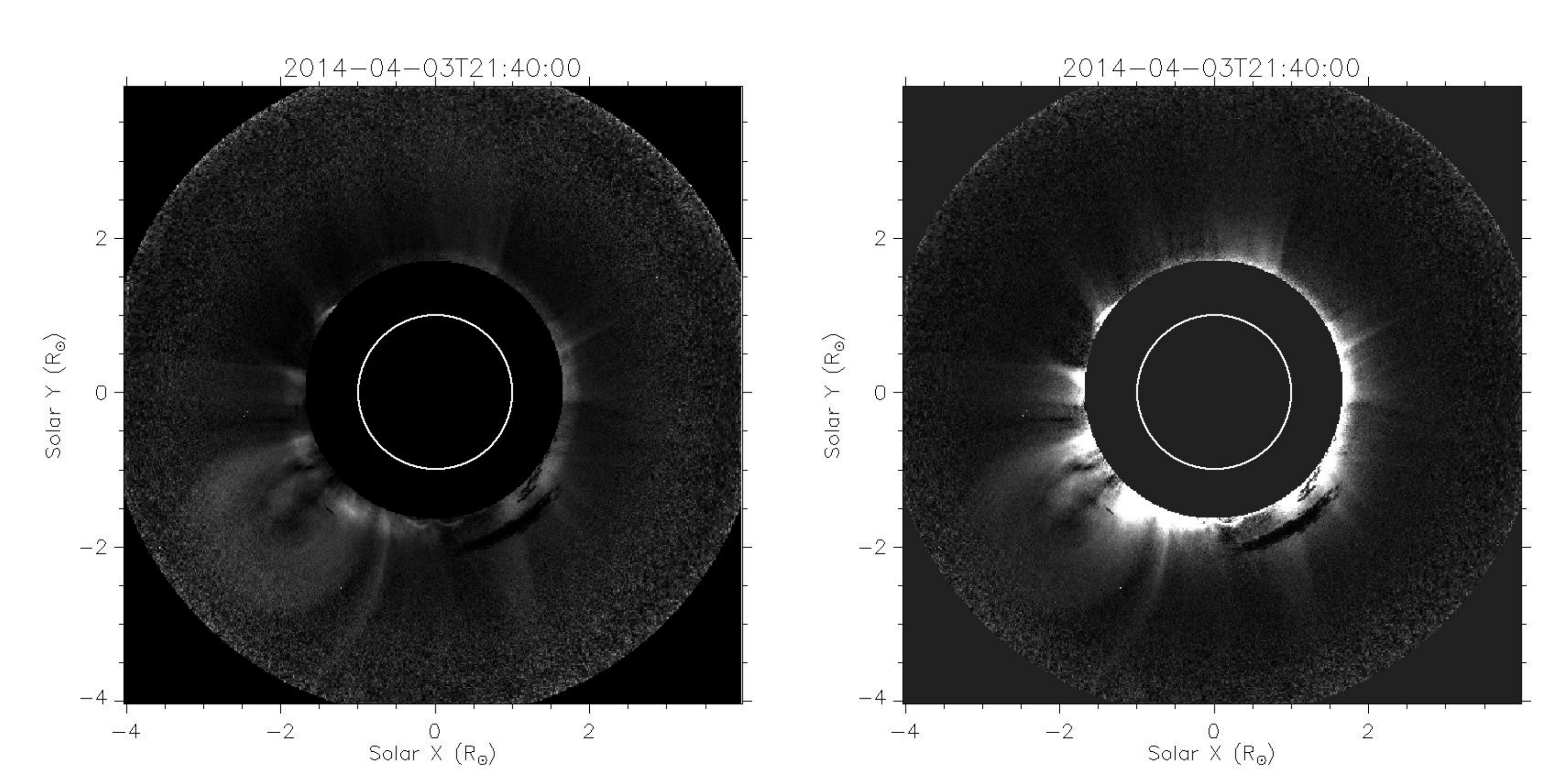}
                }
      \vspace{-0.01\textwidth}   
\caption{A comparison of application of SiRGraF and NRGF on images of STEREO/COR-1A observed at the maximum period of the solar-cycle on \edcom{03 April 2014} presented in {\it left} and {\it right} panels respectively. (An animation is available in the \edcom{Electronic Supplementary Material}.)}
   \label{fig:cor1max}
\end{figure}

We have applied SiRGraF to different datasets including LASCO-C2, STEREO/COR-1A and COR-2A, and KCor with observations covering different phases of the solar cycle. Figure \ref{fig:algo_op} shows the output of the algorithm when applied to COR-1A images near minimum of Solar Cycle 24. The result is also compared with the outcome of NRGF images in Figure \ref{fig:nrgf}. We also tested SiRGraF on COR-1A images for the observations of \edcom{03 April 2014} during the maximum phase of the solar cycle. The COR-1 images have resulted in poor SNR over the years with an uncertain jitter pattern in the images and hence degrading the image quality over Solar Cycle 24 ({see \urlurl{cor1.gsfc.nasa.gov/docs/COR1\_status.pdf}}).
Application to such images with poor SNR served as a crucial test for our algorithm. This was also compared with application of NRGF on the same set of images as shown in Figure~\ref{fig:cor1max}. The output of SiRGraF is in the left panel whereas NRGF \edcom{is shown} on the right. Both the algorithms appear to work \edcom{similarly} for the COR-1A images observed during the solar maximum. It is evident that the inner FOV appears brighter for NRGF images as compared to SiRGraF output, and \edcom{this} has already been mentioned in Section \ref{sec:nrgfcompare}. As a result, a difference could be observed in identification of the structures close to the inner FOV. The first appearance of the CME in the top panel appears to be obscured by the intensity in the NRGF-processed images whereas the same could be seen clearly for SiRGraF-filtered images. {This is an essential requirement for the study of kinematics of CMEs at the lower heights with only white-light observations. Any discrepancy in the identification of a tracked feature at the lower heights might \edcom{result} in leaving out the impulsive acceleration phase, or an erroneous measurement can lead to spurious results.} The evolution of the CME through the FOV could be seen in successive frames. It is observed that SiRGraF provides a uniform identification of different structures of the CME at all the heights, unlike NRGF. This could serve as an important application of our algorithm on COR-1 images enabling us to better observe CMEs close to the Sun. The usefulness of both of these algorithms is limited to the lower heights because SNR becomes low towards the { outer edge} of the FOV { (beyond $\approx$\,3.5\,R$_\odot$)} in COR-1 images, as evident in both cases.

\begin{figure}[!htbp]   
   \centerline{\hspace*{0.05\textwidth}
               \includegraphics[width=0.7\textwidth,clip=]{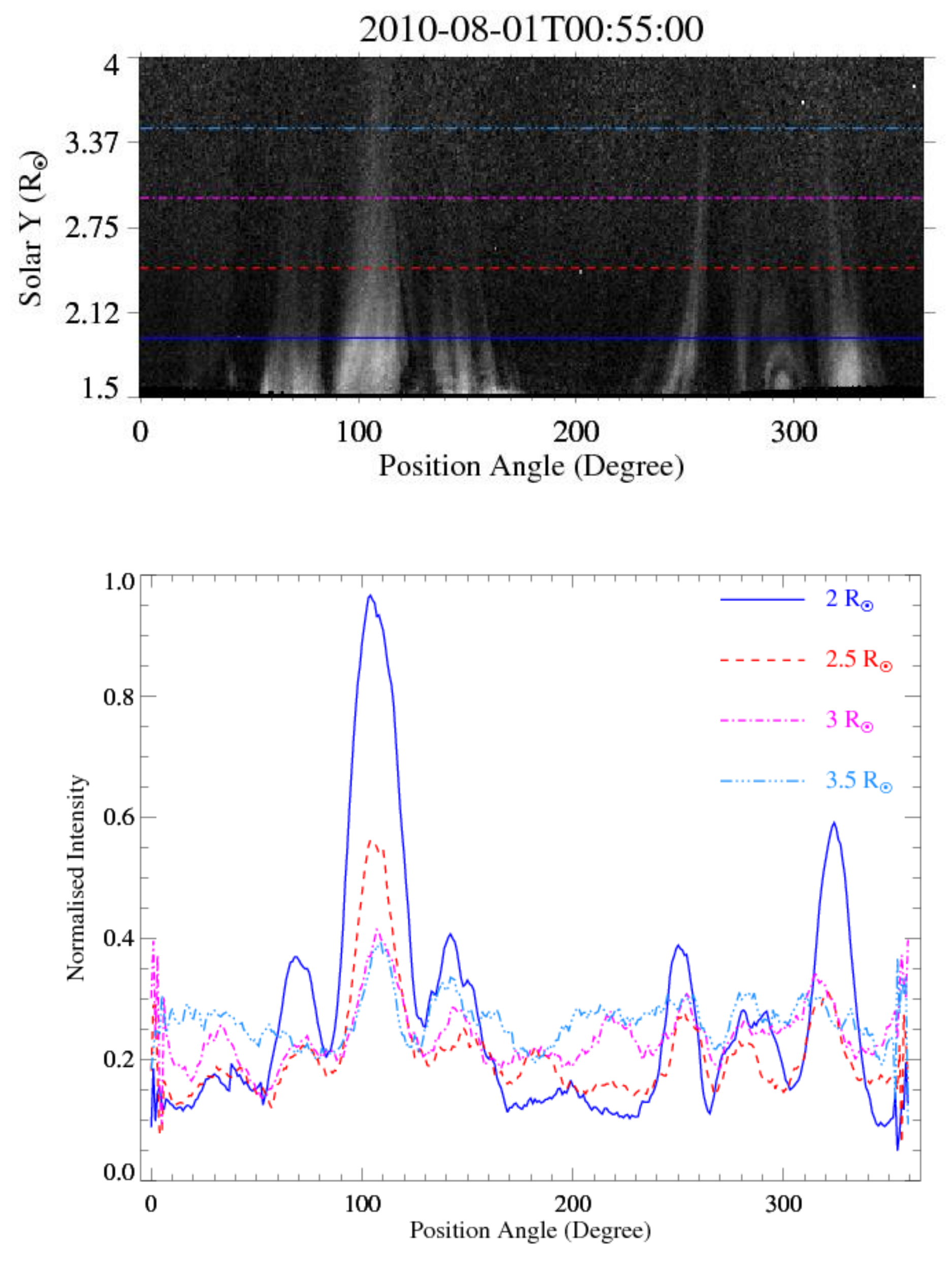}
               \includegraphics[width=0.7\textwidth,clip=]{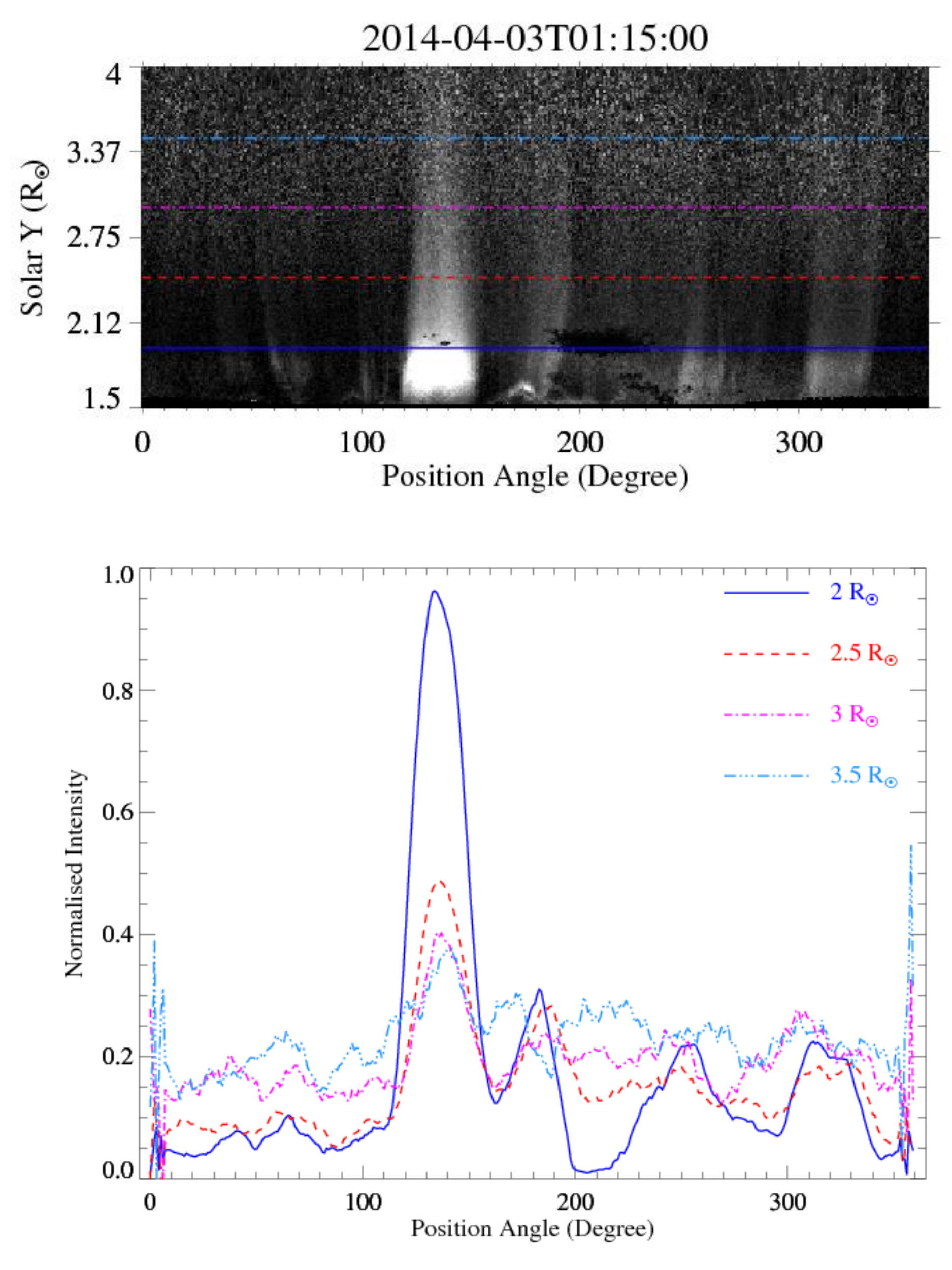}}
      \vspace{-0.01\textwidth} 
      \vspace{-0.01\textwidth}   
\caption{Radial intensity variation after the application of SiRGraF to COR-1A images taken during solar minima ({\it left}) and maxima ({\it right}). The {\it upper panel} shows the images converted to polar coordinates while the {\it lower panel} shows the normalised intensities at the four heights marked in the images of the upper panel.}
   \label{fig:radinty_comp}
\end{figure}

For the COR-1A images observed during solar maxima and minima and processed with SiRGraF, we compared the radial variation in the intensity at all the position angles (PAs) as shown in Figure \ref{fig:radinty_comp}. The top panel shows the images taken at the two phases of the solar cycle converted to polar coordinates while the bottom panel shows the normalised intensity variations at the four heights marked in the top panel. It can be seen that for both of the images the intensity in the processed images are close to each other at all the heights except at the bright streamer regions. In the intensity plot for \edcom{03 April 2014} image, there is a sharp dip in the intensity at 2\,R$_\odot$ at PA of $\approx$\,200$^\circ$. It should be noted that at this particular location there is some dark artifact in the corresponding image shown in the top panel, which is responsible for this dip. Other regions with \edcom{fewer} structures does not show a steep change in intensity from inner to outer FOV for both images. The performance of the algorithm on the images observed during different phases of the solar cycle enhances its utility for studying the coronal dynamics over long periods.

\section{Summary}
    \label{sec: summary}

To study the dynamics of the solar atmosphere using coronagraph images, it is important to visualise the structures well throughout the FOV. The radial gradient in the intensity of the corona makes it difficult to observe the structures throughout the FOV with same brightness and contrast in the images. 
To reduce the radial gradient in the intensity in such huge number of images in a faster manner and bring out the details of coronal transients, we have developed an algorithm, the {\it Simple Radial Gradient Filter} (SiRGraF). The algorithm requires two backgrounds for processing a coronagraph image. One is a minimum-background image prepared from atleast a day of images to filter out the constant background and another one is a uniform-intensity circularly symmetric background prepared using the \edcom{azimuthal} average radial-intensity profile from the minimum background. The minimum background is subtracted from individual images followed by division of the difference image by the uniform background. The images processed with this filter appear to show identification of the coronal structures throughout the coronagraph FOV with uniform intensity. {We found that this algorithm works well in identifying the coronal structures in the coronagraph images over different phases of the solar cycle. These include the ones with good SNR, such as LASCO-C2, STEREO/COR-2A, and poor SNR images of STEREO/COR-1A and KCor.}
{We illustrated the effect of different periods of background used in the SiRGraF that shows for transients such as CMEs one day background perform in similar way as the extended period one. The same is not applicable while considering long lived streamers.}


{ We also compared the performance of our algorithm with NRGF and found that SiRGraF is efficient and fast when \edcom{a large number} of coronagraph images have to be processes quickly.}
The processing time of SiRGraF could be further reduced by using the minimum backgrounds which has already been generated for instruments such as STEREO/COR-1. In addition, if uniform backgrounds could also be stored in a similar way, then SiRGraF works even faster for bulk processing of images. This performance of SiRGraF is achieved maintaining the intensity contrast similar to NRGF. 
We notice in Figure \ref{fig:cor1max} that a faint halo is observed in SiRGraF processed images which is neither present in the NRGF ones nor in Figure \ref{fig:algo_op}. During the period of observation considered in the solar maxima case, STEREO/COR-1A is associated with high jitter and instrument related artifacts. Upon careful observation we found that the presence of rogue non-zero intensity values near the southwest limb outside the occulter has contributed to the background creation, which is reflected as halo in the filtered images. This does not impact the identification and tracking of CMEs in subsequent images, but it will be taken care of in a future version of the algorithm. {One needs to be careful while implementing this algorithm and generating the backgrounds for data when roll maneuvers are performed for the SOHO and STEREO spacecrafts where the FOV of the instruments are rotated.} 

This kind of processing could be used while studying the kinematics and morphology of CMEs stretched over long periods of time. One should be careful \edcom{when using} this data product to estimate the CMEs mass. Such estimations require accurate intensity measurements that get flattened out in the radial-filtering process. 
{It should be noted that among the existing white-light coronagraphs, STEREO/COR-1 observes the inner corona closest up to 1.4\,R$_\odot$ and KCor upto 1.05\,R$_\odot$.
The application of SiRGraF on large amount of COR-1 and KCor data could provide us an improvement in understanding the CME dynamics in the inner corona.} Moreover, it could also pave a path for the development of an automated CME-detection algorithm to work well with COR-1 and KCor images and generate statistics of CME properties during their early evolution. 
Processing COR-1 and KCor images with SiRGraF will also be useful as
inputs to stereoscopic forward modelling using Graduated Cylindrical
Shell \citep{Thernisien2006ApJ, Thernisien2009SoPh, Thernisien2011ApJS,
Majumdar2020ApJM}. The application of the algorithm is not only
limited to CMEs, but also to analyse the dynamics
of streamers, blobs, current sheets/rays, etc... in this FOV, and long-term
coronal studies  \citep{Lamy2020c2, Lamy2020c3}. The recently launched
{\it Multi Element Telescope for Imaging and Spectroscopy}
(METIS) onboard \textit{Solar Orbiter} \citep{2020A&A...642A...1M},
and future space-based missions including the {\it Visible Emission Line
Coronagraph} \citep[VELC: ][]{VELC17, Banerjee2017} \edcom{onboard} {\it Aditya-L1}
\citep{ADITYA2017}, and the {\it Association of Spacecraft for
Polarimetric and the Imaging Investigation of the Corona of the Sun} \citep[ASPIICS:][]{Proba3}
of the {\it Project for Onboard Autonomy-3} (PROBA-3)
will also observe the inner corona. The application of SiRGraF on their
data will be helpful to bring out processes occurring in the inner corona
thereby improving our understanding of the same.

%

\begin{acks}
We would like to thank the anonymous \edcom{reviewer} for \edcom{their} valuable suggestions, which have enabled us to improve the quality of the manuscript.
We would like to thank the IIA and ARIES for providing the required computational facilities. 
The SECCHI data used here were produced by an international consortium of the Naval Research Laboratory (USA), Lockheed Martin Solar and Astrophysics Lab (USA), NASA Goddard Space Flight Center (USA), Rutherford Appleton Laboratory (UK), University of Birmingham (UK), Max-Planck-Institut for Solar System Research (Germany), Centre Spatiale de Li\`ege (Belgium), Institut d'Optique Th\'eorique et Appliqu\'ee (France), and the Institut d'Astrophysique Spatiale (France). SOHO is a project of international cooperation between ESA and NASA. We also thank the Mauna Loa Solar Observatory, operated by the High Altitude Observatory for making KCor data available. We also thank NASA for making SOHO/LASCO data publicly available.
\end{acks}

\begin{fundinginformation}
R. Patel and S. Majumdar are supported by the Department of Science and Technology, Govt. of India for their research at the IIA and ARIES.
\end{fundinginformation}

\begin{dataavailability}
The datasets generated during the current study are available from the corresponding author on reasonable request.
\end{dataavailability}

\begin{ethics}
\begin{conflict}
The authors declare that they have no conflict of interests.
\end{conflict}
\end{ethics}

\bibliographystyle{spr-mp-sola}
\bibliography{bibli}  

\begin{thebibliography}{53}
\ifx\bisbn     \undefined \def\bisbn  #1{ISBN #1}\fi
\ifx\binits    \undefined \def\binits#1{#1}\fi
\ifx\bauthor   \undefined \def\bauthor#1{#1}\fi
\ifx\batitle   \undefined \def\batitle#1{#1}\fi
\ifx\bjtitle   \undefined \def\bjtitle#1{\textit{#1}}\fi
\ifx\bvolume   \undefined \def\bvolume#1{\textbf{#1}}\fi
\ifx\byear     \undefined \def\byear#1{#1}\fi
\ifx\bissue    \undefined \def\bissue#1{#1}\fi
\ifx\bfpage    \undefined \def\bfpage#1{#1}\fi
\ifx\blpage    \undefined \def\blpage #1{#1}\fi
\ifx\burl      \undefined \def\burl#1{#1}\fi
\ifx\href      \undefined \def\href#1#2{#2}\fi
\ifx\betal     \undefined \def\betal{et al.}\fi
\ifx\bctitle   \undefined \def\bctitle#1{#1}\fi
\ifx\beditor   \undefined \def\beditor#1{#1}\fi
\ifx\bbtitle   \undefined \def\bbtitle#1{\textit{#1}}\fi
\ifx\bedition  \undefined \def\bedition#1{#1}\fi
\ifx\bseriesno \undefined \def\bseriesno#1{\textbf{#1}}\fi
\ifx\blocation \undefined \def\blocation#1{#1}\fi
\ifx\bsertitle \undefined \def\bsertitle#1{\textit{#1}}\fi
\ifx\bsnm      \undefined \def\bsnm#1{#1}\fi
\ifx\bsuffix   \undefined \def\bsuffix#1{#1}\fi
\ifx\bparticle \undefined \def\bparticle#1{#1}\fi
\ifx\barticle  \undefined \def\barticle#1{}\fi
\ifx\binstitute  \undefined \def\binstitute#1{#1}\fi
\ifx\bpublisher  \undefined \def\bpublisher#1{#1}\fi
\ifx\doiurl    \undefined \def\doiurl#1{\href{#1}{DOI}}\fi
\makeatletter
\def\safeHref#1#2#3{\in@{http}{#2}\ifin@\href{#2}{#3}\else\href{#1#2}{#3}\fi}
\makeatother
\ifx\adsurl    \undefined
  \def\adsurl#1{\safeHref{https://ui.adsabs.harvard.edu/abs/}{#1}{ADS}}\fi
\ifx\arxivurl  \undefined
  \def\arxivurl#1{\safeHref{http://arxiv.org/abs/}{#1}{arXiv}}\fi
\ifx\botherref \undefined \def\botherref#1{}\fi
\ifx\url       \undefined \def\url#1{#1}\fi
\ifx\bchapter  \undefined \def\bchapter#1{}\fi
\ifx\bbook     \undefined \def\bbook#1{}\fi
\ifx\bcomment  \undefined \def\bcomment#1{#1}\fi
\ifx\oauthor   \undefined \def\oauthor#1{#1}\fi
\ifx\citeauthoryear \undefined\def \citeauthoryear#1{#1}\fi
\def\endbibitem {}
\ifx\bconflocation  \undefined \def\bconflocation#1{#1} \fi

\bibitem[\protect\citeauthoryear{Banerjee, Patel, and
  Pant}{2017}]{Banerjee2017}
\begin{barticle}
\bauthor{\bsnm{Banerjee}, \binits{D.}},
\bauthor{\bsnm{Patel}, \binits{R.}},
\bauthor{\bsnm{Pant}, \binits{V.}}:
\byear{2017},
\batitle{The Inner Coronagraph on Board ADITYA-L1 and Automatic Detection of
  CMEs}.
\bjtitle{Proceedings of the International Astronomical Union}
\bvolume{13},
\bfpage{340–343}.
\doiurl{https://doi.org/10.1017/S1743921317008584}.
\end{barticle}
\endbibitem

\bibitem[\protect\citeauthoryear{Baumbach}{1937}]{Baumbach37}
\begin{barticle}
\bauthor{\bsnm{Baumbach}, \binits{S.}}:
\byear{1937},
\batitle{Strahlung, Ergiebigkeit und Elektronendichte der Sonnenkorona}.
\bjtitle{Astronomische Nachrichten}
\bvolume{263},
\bfpage{121}.
\doiurl{https://doi.org/10.1002/asna.19372630602}.
\burl{https://onlinelibrary.wiley.com/doi/abs/10.1002/asna.19372630602}.
\end{barticle}
\endbibitem

\bibitem[\protect\citeauthoryear{{Boe} et~al.}{2018}]{Boe2018Freezin}
\begin{barticle}
\bauthor{\bsnm{{Boe}}, \binits{B.}},
\bauthor{\bsnm{{Habbal}}, \binits{S.}},
\bauthor{\bsnm{{Druckm{\"u}ller}}, \binits{M.}},
\bauthor{\bsnm{{Landi}}, \binits{E.}},
\bauthor{\bsnm{{Kourkchi}}, \binits{E.}},
\bauthor{\bsnm{{Ding}}, \binits{A.}},
\bauthor{\bsnm{{Starha}}, \binits{P.}},
\bauthor{\bsnm{{Hutton}}, \binits{J.}}:
\byear{2018},
\batitle{{The First Empirical Determination of the Fe$^{10+}$ and Fe$^{13+}$
  Freeze-in Distances in the Solar Corona}}.
\bjtitle{\apj}
\bvolume{859},
\bfpage{155}.
\doiurl{https://doi.org/10.3847/1538-4357/aabfb7}.
\adsurl{2018ApJ...859..155B}.
\end{barticle}
\endbibitem

\bibitem[\protect\citeauthoryear{{Brueckner} et~al.}{1995}]{Brueckner95}
\begin{barticle}
\bauthor{\bsnm{{Brueckner}}, \binits{G.E.}},
\bauthor{\bsnm{{Howard}}, \binits{R.A.}},
\bauthor{\bsnm{{Koomen}}, \binits{M.J.}},
\bauthor{\bsnm{{Korendyke}}, \binits{C.M.}},
\bauthor{\bsnm{{Michels}}, \binits{D.J.}},
\bauthor{\bsnm{{Moses}}, \binits{J.D.}},
\bauthor{\bsnm{{Socker}}, \binits{D.G.}},
\bauthor{\bsnm{{Dere}}, \binits{K.P.}},
\bauthor{\bsnm{{Lamy}}, \binits{P.L.}},
\bauthor{\bsnm{{Llebaria}}, \binits{A.}},
\bauthor{\bsnm{{Bout}}, \binits{M.V.}},
\bauthor{\bsnm{{Schwenn}}, \binits{R.}},
\bauthor{\bsnm{{Simnett}}, \binits{G.M.}},
\bauthor{\bsnm{{Bedford}}, \binits{D.K.}},
\bauthor{\bsnm{{Eyles}}, \binits{C.J.}}:
\byear{1995},
\batitle{{The Large Angle Spectroscopic Coronagraph (LASCO)}}.
\bjtitle{\solphys}
\bvolume{162},
\bfpage{357}.
\doiurl{https://doi.org/10.1007/BF00733434}.
\adsurl{1995SoPh..162..357B}.
\end{barticle}
\endbibitem

\bibitem[\protect\citeauthoryear{{Byrne} et~al.}{2009}]{Byrne2009A&A}
\begin{barticle}
\bauthor{\bsnm{{Byrne}}, \binits{J.P.}},
\bauthor{\bsnm{{Gallagher}}, \binits{P.T.}},
\bauthor{\bsnm{{McAteer}}, \binits{R.T.J.}},
\bauthor{\bsnm{{Young}}, \binits{C.A.}}:
\byear{2009},
\batitle{{The kinematics of coronal mass ejections using multiscale methods}}.
\bjtitle{\aap}
\bvolume{495},
\bfpage{325}.
\doiurl{https://doi.org/10.1051/0004-6361:200809811}.
\adsurl{2009A&A...495..325B}.
\end{barticle}
\endbibitem

\bibitem[\protect\citeauthoryear{{Byrne} et~al.}{2012}]{Byrne12}
\begin{barticle}
\bauthor{\bsnm{{Byrne}}, \binits{J.P.}},
\bauthor{\bsnm{{Morgan}}, \binits{H.}},
\bauthor{\bsnm{{Habbal}}, \binits{S.R.}},
\bauthor{\bsnm{{Gallagher}}, \binits{P.T.}}:
\byear{2012},
\batitle{{Automatic Detection and Tracking of Coronal Mass Ejections. II.
  Multiscale Filtering of Coronagraph Images}}.
\bjtitle{\apj}
\bvolume{752},
\bfpage{145}.
\doiurl{https://doi.org/10.1088/0004-637X/752/2/145}.
\adsurl{2012ApJ...752..145B}.
\end{barticle}
\endbibitem

\bibitem[\protect\citeauthoryear{{de Wijn} et~al.}{2012}]{deWijn}
\begin{bchapter}
\bauthor{\bsnm{{de Wijn}}, \binits{A.G.}},
\bauthor{\bsnm{{Burkepile}}, \binits{J.T.}},
\bauthor{\bsnm{{Tomczyk}}, \binits{S.}},
\bauthor{\bsnm{{Nelson}}, \binits{P.G.}},
\bauthor{\bsnm{{Huang}}, \binits{P.}},
\bauthor{\bsnm{{Gallagher}}, \binits{D.}}:
\byear{2012},
\bctitle{{Stray light and polarimetry considerations for the COSMO
  K-Coronagraph}}.
In: \bbtitle{Ground-based and Airborne Telescopes IV},
\bsertitle{Proceedings of SPIE}
\bseriesno{8444},
\bfpage{84443N}.
\doiurl{https://doi.org/10.1117/12.926511}.
\adsurl{2012SPIE.8444E..3ND}.
\end{bchapter}
\endbibitem

\bibitem[\protect\citeauthoryear{{DeForest}, {Howard}, and
  {McComas}}{2014}]{DeForest14}
\begin{barticle}
\bauthor{\bsnm{{DeForest}}, \binits{C.E.}},
\bauthor{\bsnm{{Howard}}, \binits{T.A.}},
\bauthor{\bsnm{{McComas}}, \binits{D.J.}}:
\byear{2014},
\batitle{{Inbound Waves in the Solar Corona: A Direct Indicator of Alfv{\'e}n
  Surface Location}}.
\bjtitle{\apj}
\bvolume{787},
\bfpage{124}.
\doiurl{https://doi.org/10.1088/0004-637X/787/2/124}.
\adsurl{2014ApJ...787..124D}.
\end{barticle}
\endbibitem

\bibitem[\protect\citeauthoryear{{DeForest} et~al.}{2018}]{DeForest2018ApJ}
\begin{barticle}
\bauthor{\bsnm{{DeForest}}, \binits{C.E.}},
\bauthor{\bsnm{{Howard}}, \binits{R.A.}},
\bauthor{\bsnm{{Velli}}, \binits{M.}},
\bauthor{\bsnm{{Viall}}, \binits{N.}},
\bauthor{\bsnm{{Vourlidas}}, \binits{A.}}:
\byear{2018},
\batitle{{The Highly Structured Outer Solar Corona}}.
\bjtitle{\apj}
\bvolume{862},
\bfpage{18}.
\doiurl{https://doi.org/10.3847/1538-4357/aac8e3}.
\adsurl{2018ApJ...862...18D}.
\end{barticle}
\endbibitem

\bibitem[\protect\citeauthoryear{{Druckm{\"u}ller}}{2013}]{NAFE2013ApJS}
\begin{barticle}
\bauthor{\bsnm{{Druckm{\"u}ller}}, \binits{M.}}:
\byear{2013},
\batitle{{A Noise Adaptive Fuzzy Equalization Method for Processing Solar
  Extreme Ultraviolet Images}}.
\bjtitle{\apjs}
\bvolume{207},
\bfpage{25}.
\doiurl{https://doi.org/10.1088/0067-0049/207/2/25}.
\adsurl{2013ApJS..207...25D}.
\end{barticle}
\endbibitem

\bibitem[\protect\citeauthoryear{{Druckm{\"u}ller}, {Ru{\v{s}}in}, and
  {Minarovjech}}{2006}]{Druckm2006}
\begin{barticle}
\bauthor{\bsnm{{Druckm{\"u}ller}}, \binits{M.}},
\bauthor{\bsnm{{Ru{\v{s}}in}}, \binits{V.}},
\bauthor{\bsnm{{Minarovjech}}, \binits{M.}}:
\byear{2006},
\batitle{{A new numerical method of total solar eclipse photography
  processing}}.
\bjtitle{Contributions of the Astronomical Observatory Skalnate Pleso}
\bvolume{36},
\bfpage{131}.
\adsurl{2006CoSka..36..131D}.
\end{barticle}
\endbibitem

\bibitem[\protect\citeauthoryear{Druckmüllerov{\'{a}}, Morgan, and
  Habbal}{2011}]{FNRGF2011}
\begin{barticle}
\bauthor{\bsnm{Druckmüllerov{\'{a}}}, \binits{H.}},
\bauthor{\bsnm{Morgan}, \binits{H.}},
\bauthor{\bsnm{Habbal}, \binits{S.R.}}:
\byear{2011},
\batitle{{ENHANCING} {CORONAL} {STRUCTURES} {WITH} {THE} {FOURIER}
  {NORMALIZING}-{RADIAL}-{GRADED} {FILTER}}.
\bjtitle{The Astrophysical Journal}
\bvolume{737},
\bfpage{88}.
\doiurl{https://doi.org/10.1088/0004-637x/737/2/88}.
\burl{https://doi.org/10.1088/0004-637x/737/2/88}.
\end{barticle}
\endbibitem

\bibitem[\protect\citeauthoryear{{Espenak}}{2000}]{Espenak2000}
\begin{bchapter}
\bauthor{\bsnm{{Espenak}}, \binits{F.}}:
\byear{2000},
\bctitle{{Digital Compositing Techniques for Coronal Imaging (Invited
  review)}}.
In: \beditor{\bsnm{{Livingston}}, \binits{W.}},
\beditor{\bsnm{{{\"O}zg{\"u}{\c{c}}}}, \binits{A.}} (eds.)
\bbtitle{Last Total Solar Eclipse of the Millennium},
\bsertitle{Astronomical Society of the Pacific Conference Series}
\bseriesno{205},
\bfpage{101}.
\adsurl{2000ASPC..205..101E}.
\end{bchapter}
\endbibitem

\bibitem[\protect\citeauthoryear{{Habbal} et~al.}{2010}]{Habbal2010ApJ}
\begin{barticle}
\bauthor{\bsnm{{Habbal}}, \binits{S.R.}},
\bauthor{\bsnm{{Druckm{\"u}ller}}, \binits{M.}},
\bauthor{\bsnm{{Morgan}}, \binits{H.}},
\bauthor{\bsnm{{Daw}}, \binits{A.}},
\bauthor{\bsnm{{Johnson}}, \binits{J.}},
\bauthor{\bsnm{{Ding}}, \binits{A.}},
\bauthor{\bsnm{{Arndt}}, \binits{M.}},
\bauthor{\bsnm{{Esser}}, \binits{R.}},
\bauthor{\bsnm{{Ru{\v{s}}in}}, \binits{V.}},
\bauthor{\bsnm{{Scholl}}, \binits{I.}}:
\byear{2010},
\batitle{{Mapping the Distribution of Electron Temperature and Fe Charge States
  in the Corona with Total Solar Eclipse Observations}}.
\bjtitle{\apj}
\bvolume{708},
\bfpage{1650}.
\doiurl{https://doi.org/10.1088/0004-637X/708/2/1650}.
\adsurl{2010ApJ...708.1650H}.
\end{barticle}
\endbibitem

\bibitem[\protect\citeauthoryear{{Habbal} et~al.}{2011}]{Habbal2011ApJ}
\begin{barticle}
\bauthor{\bsnm{{Habbal}}, \binits{S.R.}},
\bauthor{\bsnm{{Druckm{\"u}ller}}, \binits{M.}},
\bauthor{\bsnm{{Morgan}}, \binits{H.}},
\bauthor{\bsnm{{Ding}}, \binits{A.}},
\bauthor{\bsnm{{Johnson}}, \binits{J.}},
\bauthor{\bsnm{{Druckm{\"u}llerov{\'a}}}, \binits{H.}},
\bauthor{\bsnm{{Daw}}, \binits{A.}},
\bauthor{\bsnm{{Arndt}}, \binits{M.B.}},
\bauthor{\bsnm{{Dietzel}}, \binits{M.}},
\bauthor{\bsnm{{Saken}}, \binits{J.}}:
\byear{2011},
\batitle{{Thermodynamics of the Solar Corona and Evolution of the Solar
  Magnetic Field as Inferred from the Total Solar Eclipse Observations of 2010
  July 11}}.
\bjtitle{\apj}
\bvolume{734},
\bfpage{120}.
\doiurl{https://doi.org/10.1088/0004-637X/734/2/120}.
\adsurl{2011ApJ...734..120H}.
\end{barticle}
\endbibitem

\bibitem[\protect\citeauthoryear{{He} et~al.}{2009}]{He2009A&A}
\begin{barticle}
\bauthor{\bsnm{{He}}, \binits{J.-S.}},
\bauthor{\bsnm{{Tu}}, \binits{C.-Y.}},
\bauthor{\bsnm{{Marsch}}, \binits{E.}},
\bauthor{\bsnm{{Guo}}, \binits{L.-J.}},
\bauthor{\bsnm{{Yao}}, \binits{S.}},
\bauthor{\bsnm{{Tian}}, \binits{H.}}:
\byear{2009},
\batitle{{Upward propagating high-frequency Alfv{\'e}n waves as identified from
  dynamic wave-like spicules observed by SOT on Hinode}}.
\bjtitle{\aap}
\bvolume{497},
\bfpage{525}.
\doiurl{https://doi.org/10.1051/0004-6361/200810777}.
\adsurl{2009A&A...497..525H}.
\end{barticle}
\endbibitem

\bibitem[\protect\citeauthoryear{{Howard} et~al.}{2008}]{Howard2008SSRv}
\begin{barticle}
\bauthor{\bsnm{{Howard}}, \binits{R.A.}},
\bauthor{\bsnm{{Moses}}, \binits{J.D.}},
\bauthor{\bsnm{{Vourlidas}}, \binits{A.}},
\bauthor{\bsnm{{Newmark}}, \binits{J.S.}},
\bauthor{\bsnm{{Socker}}, \binits{D.G.}},
\bauthor{\bsnm{{Plunkett}}, \binits{S.P.}},
\bauthor{\bsnm{{Korendyke}}, \binits{C.M.}},
\bauthor{\bsnm{{Cook}}, \binits{J.W.}},
\bauthor{\bsnm{{Hurley}}, \binits{A.}},
\bauthor{\bsnm{{Davila}}, \binits{J.M.}},
\bauthor{\bsnm{{Thompson}}, \binits{W.T.}},
\bauthor{\bsnm{{St Cyr}}, \binits{O.C.}},
\bauthor{\bsnm{{Mentzell}}, \binits{E.}},
\bauthor{\bsnm{{Mehalick}}, \binits{K.}},
\bauthor{\bsnm{{Lemen}}, \binits{J.R.}},
\bauthor{\bsnm{{Wuelser}}, \binits{J.P.}},
\bauthor{\bsnm{{Duncan}}, \binits{D.W.}},
\bauthor{\bsnm{{Tarbell}}, \binits{T.D.}},
\bauthor{\bsnm{{Wolfson}}, \binits{C.J.}},
\bauthor{\bsnm{{Moore}}, \binits{A.}},
\bauthor{\bsnm{{Harrison}}, \binits{R.A.}},
\bauthor{\bsnm{{Waltham}}, \binits{N.R.}},
\bauthor{\bsnm{{Lang}}, \binits{J.}},
\bauthor{\bsnm{{Davis}}, \binits{C.J.}},
\bauthor{\bsnm{{Eyles}}, \binits{C.J.}},
\bauthor{\bsnm{{Mapson-Menard}}, \binits{H.}},
\bauthor{\bsnm{{Simnett}}, \binits{G.M.}},
\bauthor{\bsnm{{Halain}}, \binits{J.P.}},
\bauthor{\bsnm{{Defise}}, \binits{J.M.}},
\bauthor{\bsnm{{Mazy}}, \binits{E.}},
\bauthor{\bsnm{{Rochus}}, \binits{P.}},
\bauthor{\bsnm{{Mercier}}, \binits{R.}},
\bauthor{\bsnm{{Ravet}}, \binits{M.F.}},
\bauthor{\bsnm{{Delmotte}}, \binits{F.}},
\bauthor{\bsnm{{Auchere}}, \binits{F.}},
\bauthor{\bsnm{{Delaboudiniere}}, \binits{J.P.}},
\bauthor{\bsnm{{Bothmer}}, \binits{V.}},
\bauthor{\bsnm{{Deutsch}}, \binits{W.}},
\bauthor{\bsnm{{Wang}}, \binits{D.}},
\bauthor{\bsnm{{Rich}}, \binits{N.}},
\bauthor{\bsnm{{Cooper}}, \binits{S.}},
\bauthor{\bsnm{{Stephens}}, \binits{V.}},
\bauthor{\bsnm{{Maahs}}, \binits{G.}},
\bauthor{\bsnm{{Baugh}}, \binits{R.}},
\bauthor{\bsnm{{McMullin}}, \binits{D.}},
\bauthor{\bsnm{{Carter}}, \binits{T.}}:
\byear{2008},
\batitle{{Sun Earth Connection Coronal and Heliospheric Investigation
  (SECCHI)}}.
\bjtitle{\ssr}
\bvolume{136},
\bfpage{67}.
\doiurl{https://doi.org/10.1007/s11214-008-9341-4}.
\adsurl{2008SSRv..136...67H}.
\end{barticle}
\endbibitem

\bibitem[\protect\citeauthoryear{{Howard}}{2011}]{Howard11}
\begin{bbook}
\bauthor{\bsnm{{Howard}}, \binits{T.}}:
\byear{2011},
\bbtitle{{Coronal Mass Ejections: An Introduction}}
\bseriesno{376}.
\doiurl{https://doi.org/10.1007/978-1-4419-8789-1}.
\adsurl{2011ASSL..376.....H}.
\end{bbook}
\endbibitem

\bibitem[\protect\citeauthoryear{{Hulst}}{1950}]{Hulst50}
\begin{barticle}
\bauthor{\bsnm{{Hulst}}, \binits{d.} \bsuffix{van}}:
\byear{1950},
\batitle{{The electron density of the solar corona}}.
\bjtitle{Bulletin of the Astronomical Institutes of the Netherlands}
\bvolume{11},
\bfpage{135}.
\end{barticle}
\endbibitem

\bibitem[\protect\citeauthoryear{{Hutton} and {Morgan}}{2017}]{ACT2017A&A}
\begin{barticle}
\bauthor{\bsnm{{Hutton}}, \binits{J.}},
\bauthor{\bsnm{{Morgan}}, \binits{H.}}:
\byear{2017},
\batitle{{Automated detection of coronal mass ejections in three-dimensions
  using multi-viewpoint observations}}.
\bjtitle{\aap}
\bvolume{599},
\bfpage{A68}.
\doiurl{https://doi.org/10.1051/0004-6361/201629516}.
\adsurl{2017A&A...599A..68H}.
\end{barticle}
\endbibitem

\bibitem[\protect\citeauthoryear{{Koutchmy} et~al.}{1992}]{Koutchmy1992A&A}
\begin{barticle}
\bauthor{\bsnm{{Koutchmy}}, \binits{S.}},
\bauthor{\bsnm{{Altrock}}, \binits{R.C.}},
\bauthor{\bsnm{{Darvann}}, \binits{T.A.}},
\bauthor{\bsnm{{Dzubenko}}, \binits{N.I.}},
\bauthor{\bsnm{{Henry}}, \binits{T.W.}},
\bauthor{\bsnm{{Kim}}, \binits{I.}},
\bauthor{\bsnm{{Koutchmy}}, \binits{O.}},
\bauthor{\bsnm{{Martinez}}, \binits{P.}},
\bauthor{\bsnm{{Nitschelm}}, \binits{C.}},
\bauthor{\bsnm{{Rubo}}, \binits{G.A.}}:
\byear{1992},
\batitle{{Coronal photometry and analysis of the eclipse corona of July 22,
  1990}}.
\bjtitle{\aaps}
\bvolume{96},
\bfpage{169}.
\adsurl{1992A&AS...96..169K}.
\end{barticle}
\endbibitem

\bibitem[\protect\citeauthoryear{{Lamy} et~al.}{2020a}]{Lamy2020c2}
\begin{barticle}
\bauthor{\bsnm{{Lamy}}, \binits{P.}},
\bauthor{\bsnm{{Llebaria}}, \binits{A.}},
\bauthor{\bsnm{{Boclet}}, \binits{B.}},
\bauthor{\bsnm{{Gilardy}}, \binits{H.}},
\bauthor{\bsnm{{Burtin}}, \binits{M.}},
\bauthor{\bsnm{{Floyd}}, \binits{O.}}:
\byear{2020}a,
\batitle{{Coronal Photopolarimetry with the LASCO-C2 Coronagraph over 24 Years
  [1996 - 2019]}}.
\bjtitle{\solphys}
\bvolume{295},
\bfpage{89}.
\doiurl{https://doi.org/10.1007/s11207-020-01650-y}.
\adsurl{2020SoPh..295...89L}.
\end{barticle}
\endbibitem

\bibitem[\protect\citeauthoryear{{Lamy} et~al.}{2020b}]{Lamy2020c3}
\begin{botherref}
\oauthor{\bsnm{{Lamy}}, \binits{P.}},
\oauthor{\bsnm{{Gilardy}}, \binits{H.}},
\oauthor{\bsnm{{Llebaria}}, \binits{A.}},
\oauthor{\bsnm{{Quemerais}}, \binits{E.}},
\oauthor{\bsnm{{Ernandes}}, \binits{F.}}:
2020b,
{Coronal Photopolarimetry with the LASCO-C3 Coronagraph over 24 Years
  [1996-2019] -- Application to the K/F Separation and to the Determination of
  the Electron Density}.
\textit{arXiv e-prints},
arXiv:2009.04820.
\adsurl{2020arXiv200904820L}.
\end{botherref}
\endbibitem

\bibitem[\protect\citeauthoryear{{Lee} et~al.}{2020}]{Lee2020ApJ}
\begin{barticle}
\bauthor{\bsnm{{Lee}}, \binits{J.-O.}},
\bauthor{\bsnm{{Cho}}, \binits{K.-S.}},
\bauthor{\bsnm{{Lee}}, \binits{K.-S.}},
\bauthor{\bsnm{{Cho}}, \binits{I.-H.}},
\bauthor{\bsnm{{Lee}}, \binits{J.}},
\bauthor{\bsnm{{Miyashita}}, \binits{Y.}},
\bauthor{\bsnm{{Kim}}, \binits{Y.-H.}},
\bauthor{\bsnm{{Kim}}, \binits{R.-S.}},
\bauthor{\bsnm{{Jang}}, \binits{S.}}:
\byear{2020},
\batitle{{Formation of Post-CME Blobs Observed by LASCO-C2 and K-Cor on 2017
  September 10}}.
\bjtitle{\apj}
\bvolume{892},
\bfpage{129}.
\doiurl{https://doi.org/10.3847/1538-4357/ab799a}.
\adsurl{2020ApJ...892..129L}.
\end{barticle}
\endbibitem

\bibitem[\protect\citeauthoryear{{Lemen} et~al.}{2012}]{AIA}
\begin{barticle}
\bauthor{\bsnm{{Lemen}}, \binits{J.R.}},
\bauthor{\bsnm{{Title}}, \binits{A.M.}},
\bauthor{\bsnm{{Akin}}, \binits{D.J.}},
\bauthor{\bsnm{{Boerner}}, \binits{P.F.}},
\bauthor{\bsnm{{Chou}}, \binits{C.}},
\bauthor{\bsnm{{Drake}}, \binits{J.F.}},
\bauthor{\bsnm{{Duncan}}, \binits{D.W.}},
\bauthor{\bsnm{{Edwards}}, \binits{C.G.}},
\bauthor{\bsnm{{Friedlaender}}, \binits{F.M.}},
\bauthor{\bsnm{{Heyman}}, \binits{G.F.}},
\bauthor{\bsnm{{Hurlburt}}, \binits{N.E.}},
\bauthor{\bsnm{{Katz}}, \binits{N.L.}},
\bauthor{\bsnm{{Kushner}}, \binits{G.D.}},
\bauthor{\bsnm{{Levay}}, \binits{M.}},
\bauthor{\bsnm{{Lindgren}}, \binits{R.W.}},
\bauthor{\bsnm{{Mathur}}, \binits{D.P.}},
\bauthor{\bsnm{{McFeaters}}, \binits{E.L.}},
\bauthor{\bsnm{{Mitchell}}, \binits{S.}},
\bauthor{\bsnm{{Rehse}}, \binits{R.A.}},
\bauthor{\bsnm{{Schrijver}}, \binits{C.J.}},
\bauthor{\bsnm{{Springer}}, \binits{L.A.}},
\bauthor{\bsnm{{Stern}}, \binits{R.A.}},
\bauthor{\bsnm{{Tarbell}}, \binits{T.D.}},
\bauthor{\bsnm{{Wuelser}}, \binits{J.-P.}},
\bauthor{\bsnm{{Wolfson}}, \binits{C.J.}},
\bauthor{\bsnm{{Yanari}}, \binits{C.}},
\bauthor{\bsnm{{Bookbinder}}, \binits{J.A.}},
\bauthor{\bsnm{{Cheimets}}, \binits{P.N.}},
\bauthor{\bsnm{{Caldwell}}, \binits{D.}},
\bauthor{\bsnm{{Deluca}}, \binits{E.E.}},
\bauthor{\bsnm{{Gates}}, \binits{R.}},
\bauthor{\bsnm{{Golub}}, \binits{L.}},
\bauthor{\bsnm{{Park}}, \binits{S.}},
\bauthor{\bsnm{{Podgorski}}, \binits{W.A.}},
\bauthor{\bsnm{{Bush}}, \binits{R.I.}},
\bauthor{\bsnm{{Scherrer}}, \binits{P.H.}},
\bauthor{\bsnm{{Gummin}}, \binits{M.A.}},
\bauthor{\bsnm{{Smith}}, \binits{P.}},
\bauthor{\bsnm{{Auker}}, \binits{G.}},
\bauthor{\bsnm{{Jerram}}, \binits{P.}},
\bauthor{\bsnm{{Pool}}, \binits{P.}},
\bauthor{\bsnm{{Soufli}}, \binits{R.}},
\bauthor{\bsnm{{Windt}}, \binits{D.L.}},
\bauthor{\bsnm{{Beardsley}}, \binits{S.}},
\bauthor{\bsnm{{Clapp}}, \binits{M.}},
\bauthor{\bsnm{{Lang}}, \binits{J.}},
\bauthor{\bsnm{{Waltham}}, \binits{N.}}:
\byear{2012},
\batitle{{The Atmospheric Imaging Assembly (AIA) on the Solar Dynamics
  Observatory (SDO)}}.
\bjtitle{\solphys}
\bvolume{275},
\bfpage{17}.
\doiurl{https://doi.org/10.1007/s11207-011-9776-8}.
\adsurl{2012SoPh..275...17L}.
\end{barticle}
\endbibitem

\bibitem[\protect\citeauthoryear{{Majumdar} et~al.}{2020}]{Majumdar2020ApJM}
\begin{barticle}
\bauthor{\bsnm{{Majumdar}}, \binits{S.}},
\bauthor{\bsnm{{Pant}}, \binits{V.}},
\bauthor{\bsnm{{Patel}}, \binits{R.}},
\bauthor{\bsnm{{Banerjee}}, \binits{D.}}:
\byear{2020},
\batitle{{Connecting 3D Evolution of Coronal Mass Ejections to Their Source
  Regions}}.
\bjtitle{\apj}
\bvolume{899},
\bfpage{6}.
\doiurl{https://doi.org/10.3847/1538-4357/aba1f2}.
\adsurl{2020ApJ...899....6M}.
\end{barticle}
\endbibitem

\bibitem[\protect\citeauthoryear{{Masson} et~al.}{2014}]{Masson2014ApJ}
\begin{barticle}
\bauthor{\bsnm{{Masson}}, \binits{S.}},
\bauthor{\bsnm{{McCauley}}, \binits{P.}},
\bauthor{\bsnm{{Golub}}, \binits{L.}},
\bauthor{\bsnm{{Reeves}}, \binits{K.K.}},
\bauthor{\bsnm{{DeLuca}}, \binits{E.E.}}:
\byear{2014},
\batitle{{Dynamics of the Transition Corona}}.
\bjtitle{\apj}
\bvolume{787},
\bfpage{145}.
\doiurl{https://doi.org/10.1088/0004-637X/787/2/145}.
\adsurl{2014ApJ...787..145M}.
\end{barticle}
\endbibitem

\bibitem[\protect\citeauthoryear{Michelson}{1927}]{michelson1927studies}
\begin{bbook}
\bauthor{\bsnm{Michelson}, \binits{A.A.}}:
\byear{1927},
\bbtitle{Studies in Optics},
\bsertitle{The Univ. of Chicago Science Series},
\bpublisher{University Press}.
\bisbn{9780226523880}.
\burl{https://books.google.co.in/books?id=FXazQgAACAAJ}.
\end{bbook}
\endbibitem

\bibitem[\protect\citeauthoryear{Morgan}{2015}]{Morgan2015}
\begin{barticle}
\bauthor{\bsnm{Morgan}, \binits{H.}}:
\byear{2015},
\batitle{{AN} {ATLAS} {OF} {CORONAL} {ELECTRON} {DENSITY} {AT} 5 R $\odot$ . I.
  {DATA} {PROCESSING} {AND} {CALIBRATION}}.
\bjtitle{The Astrophysical Journal Supplement Series}
\bvolume{219},
\bfpage{23}.
\doiurl{https://doi.org/10.1088/0067-0049/219/2/23}.
\burl{https://doi.org/10.1088/0067-0049/219/2/23}.
\end{barticle}
\endbibitem

\bibitem[\protect\citeauthoryear{{Morgan} and
  {Druckm{\"u}ller}}{2014}]{Morgan2014SoPh}
\begin{barticle}
\bauthor{\bsnm{{Morgan}}, \binits{H.}},
\bauthor{\bsnm{{Druckm{\"u}ller}}, \binits{M.}}:
\byear{2014},
\batitle{{Multi-Scale Gaussian Normalization for Solar Image Processing}}.
\bjtitle{\solphys}
\bvolume{289},
\bfpage{2945}.
\doiurl{https://doi.org/10.1007/s11207-014-0523-9}.
\adsurl{2014SoPh..289.2945M}.
\end{barticle}
\endbibitem

\bibitem[\protect\citeauthoryear{{Morgan} and {Habbal}}{2010}]{Morgan2010ApJ}
\begin{barticle}
\bauthor{\bsnm{{Morgan}}, \binits{H.}},
\bauthor{\bsnm{{Habbal}}, \binits{S.}}:
\byear{2010},
\batitle{{A Method for Separating Coronal Mass Ejections from the Quiescent
  Corona}}.
\bjtitle{\apj}
\bvolume{711},
\bfpage{631}.
\doiurl{https://doi.org/10.1088/0004-637X/711/2/631}.
\adsurl{2010ApJ...711..631M}.
\end{barticle}
\endbibitem

\bibitem[\protect\citeauthoryear{{Morgan} and {Habbal}}{2007}]{Morgan07}
\begin{barticle}
\bauthor{\bsnm{{Morgan}}, \binits{H.}},
\bauthor{\bsnm{{Habbal}}, \binits{S.R.}}:
\byear{2007},
\batitle{{The long-term stability of the visible F corona at heights of 3 -–
  6 R\textsubscript{$\odot$}}}.
\bjtitle{Astronomy \& Astrophysics}
\bvolume{471},
\bfpage{L47}.
\end{barticle}
\endbibitem

\bibitem[\protect\citeauthoryear{{Morgan}, {Byrne}, and
  {Habbal}}{2012}]{Morgan12}
\begin{barticle}
\bauthor{\bsnm{{Morgan}}, \binits{H.}},
\bauthor{\bsnm{{Byrne}}, \binits{J.P.}},
\bauthor{\bsnm{{Habbal}}, \binits{S.R.}}:
\byear{2012},
\batitle{{Automatically Detecting and Tracking Coronal Mass Ejections. I.
  Separation of Dynamic and Quiescent Components in Coronagraph Images}}.
\bjtitle{The Astrophysical Journal}
\bvolume{752},
\bfpage{144}.
\end{barticle}
\endbibitem

\bibitem[\protect\citeauthoryear{{Morgan}, {Habbal}, and
  {Woo}}{2006}]{NRGF2006SoPh}
\begin{barticle}
\bauthor{\bsnm{{Morgan}}, \binits{H.}},
\bauthor{\bsnm{{Habbal}}, \binits{S.R.}},
\bauthor{\bsnm{{Woo}}, \binits{R.}}:
\byear{2006},
\batitle{{The Depiction of Coronal Structure in White-Light Images}}.
\bjtitle{\solphys}
\bvolume{236},
\bfpage{263}.
\doiurl{https://doi.org/10.1007/s11207-006-0113-6}.
\adsurl{2006SoPh..236..263M}.
\end{barticle}
\endbibitem

\bibitem[\protect\citeauthoryear{{Morrill} et~al.}{2006}]{Morrill2006SoPh}
\begin{barticle}
\bauthor{\bsnm{{Morrill}}, \binits{J.S.}},
\bauthor{\bsnm{{Korendyke}}, \binits{C.M.}},
\bauthor{\bsnm{{Brueckner}}, \binits{G.E.}},
\bauthor{\bsnm{{Giovane}}, \binits{F.}},
\bauthor{\bsnm{{Howard}}, \binits{R.A.}},
\bauthor{\bsnm{{Koomen}}, \binits{M.}},
\bauthor{\bsnm{{Moses}}, \binits{D.}},
\bauthor{\bsnm{{Plunkett}}, \binits{S.P.}},
\bauthor{\bsnm{{Vourlidas}}, \binits{A.}},
\bauthor{\bsnm{{Esfandiari}}, \binits{E.}},
\bauthor{\bsnm{{Rich}}, \binits{N.}},
\bauthor{\bsnm{{Wang}}, \binits{D.}},
\bauthor{\bsnm{{Thernisien}}, \binits{A.F.}},
\bauthor{\bsnm{{Lamy}}, \binits{P.}},
\bauthor{\bsnm{{Llebaria}}, \binits{A.}},
\bauthor{\bsnm{{Biesecker}}, \binits{D.}},
\bauthor{\bsnm{{Michels}}, \binits{D.}},
\bauthor{\bsnm{{Gong}}, \binits{Q.}},
\bauthor{\bsnm{{Andrews}}, \binits{M.}}:
\byear{2006},
\batitle{{Calibration of the Soho/Lasco C3 White Light Coronagraph}}.
\bjtitle{\solphys}
\bvolume{233},
\bfpage{331}.
\doiurl{https://doi.org/10.1007/s11207-006-2058-1}.
\adsurl{2006SoPh..233..331M}.
\end{barticle}
\endbibitem

\bibitem[\protect\citeauthoryear{{M{\"u}ller}
  et~al.}{2020}]{2020A&A...642A...1M}
\begin{barticle}
\bauthor{\bsnm{{M{\"u}ller}}, \binits{D.}},
\bauthor{\bsnm{{St. Cyr}}, \binits{O.C.}},
\bauthor{\bsnm{{Zouganelis}}, \binits{I.}},
\bauthor{\bsnm{{Gilbert}}, \binits{H.R.}},
\bauthor{\bsnm{{Marsden}}, \binits{R.}},
\bauthor{\bsnm{{Nieves-Chinchilla}}, \binits{T.}},
\bauthor{\bsnm{{Antonucci}}, \binits{E.}},
\bauthor{\bsnm{{Auch{\`e}re}}, \binits{F.}},
\bauthor{\bsnm{{Berghmans}}, \binits{D.}},
\bauthor{\bsnm{{Horbury}}, \binits{T.S.}},
\bauthor{\bsnm{{Howard}}, \binits{R.A.}},
\bauthor{\bsnm{{Krucker}}, \binits{S.}},
\bauthor{\bsnm{{Maksimovic}}, \binits{M.}},
\bauthor{\bsnm{{Owen}}, \binits{C.J.}},
\bauthor{\bsnm{{Rochus}}, \binits{P.}},
\bauthor{\bsnm{{Rodriguez-Pacheco}}, \binits{J.}},
\bauthor{\bsnm{{Romoli}}, \binits{M.}},
\bauthor{\bsnm{{Solanki}}, \binits{S.K.}},
\bauthor{\bsnm{{Bruno}}, \binits{R.}},
\bauthor{\bsnm{{Carlsson}}, \binits{M.}},
\bauthor{\bsnm{{Fludra}}, \binits{A.}},
\bauthor{\bsnm{{Harra}}, \binits{L.}},
\bauthor{\bsnm{{Hassler}}, \binits{D.M.}},
\bauthor{\bsnm{{Livi}}, \binits{S.}},
\bauthor{\bsnm{{Louarn}}, \binits{P.}},
\bauthor{\bsnm{{Peter}}, \binits{H.}},
\bauthor{\bsnm{{Sch{\"u}hle}}, \binits{U.}},
\bauthor{\bsnm{{Teriaca}}, \binits{L.}},
\bauthor{\bsnm{{del Toro Iniesta}}, \binits{J.C.}},
\bauthor{\bsnm{{Wimmer-Schweingruber}}, \binits{R.F.}},
\bauthor{\bsnm{{Marsch}}, \binits{E.}},
\bauthor{\bsnm{{Velli}}, \binits{M.}},
\bauthor{\bsnm{{De Groof}}, \binits{A.}},
\bauthor{\bsnm{{Walsh}}, \binits{A.}},
\bauthor{\bsnm{{Williams}}, \binits{D.}}:
\byear{2020},
\batitle{{The Solar Orbiter mission. Science overview}}.
\bjtitle{\aap}
\bvolume{642},
\bfpage{A1}.
\doiurl{https://doi.org/10.1051/0004-6361/202038467}.
\adsurl{2020A&A...642A...1M}.
\end{barticle}
\endbibitem

\bibitem[\protect\citeauthoryear{{Newkirk} and
  {Harvey}}{1968}]{Newkirk1968SoPh}
\begin{barticle}
\bauthor{\bsnm{{Newkirk}}, \binits{J.} \bsuffix{Gordon}},
\bauthor{\bsnm{{Harvey}}, \binits{J.}}:
\byear{1968},
\batitle{{Coronal Polar Plumes}}.
\bjtitle{\solphys}
\bvolume{3},
\bfpage{321}.
\doiurl{https://doi.org/10.1007/BF00155166}.
\adsurl{1968SoPh....3..321N}.
\end{barticle}
\endbibitem

\bibitem[\protect\citeauthoryear{{Owaki} and {Saito}}{1967}]{Owaki1967}
\begin{barticle}
\bauthor{\bsnm{{Owaki}}, \binits{N.}},
\bauthor{\bsnm{{Saito}}, \binits{K.}}:
\byear{1967},
\batitle{{Photometry of the Solar Corona at the 1962 February Eclipse}}.
\bjtitle{\pasj}
\bvolume{19},
\bfpage{279}.
\adsurl{1967PASJ...19..279O}.
\end{barticle}
\endbibitem

\bibitem[\protect\citeauthoryear{Pasachoff et~al.}{2007}]{Pasachoff_2007}
\begin{barticle}
\bauthor{\bsnm{Pasachoff}, \binits{J.M.}},
\bauthor{\bsnm{Ru{\v{s}}in}, \binits{V.}},
\bauthor{\bsnm{Druckmuller}, \binits{M.}},
\bauthor{\bsnm{Saniga}, \binits{M.}}:
\byear{2007},
\batitle{Fine Structures in the White-Light Solar Corona at the 2006 Eclipse}.
\bjtitle{The Astrophysical Journal}
\bvolume{665},
\bfpage{824}.
\doiurl{https://doi.org/10.1086/519680}.
\burl{https://doi.org/10.1086/519680}.
\end{barticle}
\endbibitem

\bibitem[\protect\citeauthoryear{Patel et~al.}{2018}]{Patel2018}
\begin{barticle}
\bauthor{\bsnm{Patel}, \binits{R.}},
\bauthor{\bsnm{K}, \binits{A.}},
\bauthor{\bsnm{Pant}, \binits{V.}},
\bauthor{\bsnm{Banerjee}, \binits{D.}},
\bauthor{\bsnm{K.}, \binits{S.}},
\bauthor{\bsnm{Kumar}, \binits{A.}}:
\byear{2018},
\batitle{Onboard Automated CME Detection Algorithm for the Visible Emission
  Line Coronagraph on ADITYA-L1}.
\bjtitle{Solar Physics}
\bvolume{293},
\bfpage{103}.
\doiurl{https://doi.org/10.1007/s11207-018-1323-4}.
\burl{https://doi.org/10.1007/s11207-018-1323-4}.
\end{barticle}
\endbibitem

\bibitem[\protect\citeauthoryear{{Patel} et~al.}{2020}]{patel2020A&A}
\begin{barticle}
\bauthor{\bsnm{{Patel}}, \binits{R.}},
\bauthor{\bsnm{{Pant}}, \binits{V.}},
\bauthor{\bsnm{{Chandrashekhar}}, \binits{K.}},
\bauthor{\bsnm{{Banerjee}}, \binits{D.}}:
\byear{2020},
\batitle{{A statistical study of plasmoids associated with a post-CME current
  sheet}}.
\bjtitle{\aap}
\bvolume{644},
\bfpage{A158}.
\doiurl{https://doi.org/10.1051/0004-6361/202039000}.
\adsurl{2020A&A...644A.158P}.
\end{barticle}
\endbibitem

\bibitem[\protect\citeauthoryear{{Patel} et~al.}{2021}]{ciisco2020}
\begin{barticle}
\bauthor{\bsnm{{Patel}}, \binits{R.}},
\bauthor{\bsnm{{Pant}}, \binits{V.}},
\bauthor{\bsnm{{Iyer}}, \binits{P.}},
\bauthor{\bsnm{{Banerjee}}, \binits{D.}},
\bauthor{\bsnm{{Mierla}}, \binits{M.}},
\bauthor{\bsnm{{West}}, \binits{M.J.}}:
\byear{2021},
\batitle{{Automated Detection of Accelerating Solar Eruptions Using Parabolic
  Hough Transform}}.
\bjtitle{\solphys}
\bvolume{296},
\bfpage{31}.
\doiurl{https://doi.org/10.1007/s11207-021-01770-z}.
\adsurl{2021SoPh..296...31P}.
\end{barticle}
\endbibitem

\bibitem[\protect\citeauthoryear{Qiang et~al.}{2020}]{RLMF2020101383}
\begin{barticle}
\bauthor{\bsnm{Qiang}, \binits{Z.}},
\bauthor{\bsnm{Bai}, \binits{X.}},
\bauthor{\bsnm{Ji}, \binits{K.}},
\bauthor{\bsnm{Liu}, \binits{H.}},
\bauthor{\bsnm{Shang}, \binits{Z.}}:
\byear{2020},
\batitle{Enhancing coronal structures with radial local multi-scale filter}.
\bjtitle{New Astronomy}
\bvolume{79},
\bfpage{101383}.
\doiurl{https://doi.org/10.1016/j.newast.2020.101383}.
\burl{http://www.sciencedirect.com/science/article/pii/S1384107619301927}.
\end{barticle}
\endbibitem

\bibitem[\protect\citeauthoryear{{Raghavendra Prasad} et~al.}{2017}]{VELC17}
\begin{barticle}
\bauthor{\bsnm{{Raghavendra Prasad}}, \binits{B.}},
\bauthor{\bsnm{Banerjee}, \binits{D.}},
\bauthor{\bsnm{Singh}, \binits{J.}},
\bauthor{\bsnm{Nagabhushana}, \binits{S.}},
\bauthor{\bsnm{Kumar}, \binits{A.}},
\bauthor{\bsnm{Kamath}, \binits{P.U.}},
\bauthor{\bsnm{Kathiravan}, \binits{S.}},
\bauthor{\bsnm{Venkata}, \binits{S.}},
\bauthor{\bsnm{Rajkumar}, \binits{N.}},
\bauthor{\bsnm{Natarajan}, \binits{V.}},
\bauthor{\bsnm{Juneja}, \binits{M.}},
\bauthor{\bsnm{Somu}, \binits{P.}},
\bauthor{\bsnm{Pant}, \binits{V.}},
\bauthor{\bsnm{Shaji}, \binits{N.}},
\bauthor{\bsnm{Sankarsubramanian}, \binits{K.}},
\bauthor{\bsnm{Patra}, \binits{A.}},
\bauthor{\bsnm{Venkateswaran}, \binits{R.}},
\bauthor{\bsnm{Adoni}, \binits{A.A.}},
\bauthor{\bsnm{Narendra}, \binits{S.}},
\bauthor{\bsnm{Haridas}, \binits{T.R.}},
\bauthor{\bsnm{Mathew}, \binits{S.K.}},
\bauthor{\bsnm{Krishna}, \binits{R.M.}},
\bauthor{\bsnm{Amareswari}, \binits{K.}},
\bauthor{\bsnm{Jaiswal}, \binits{B.}}:
\byear{2017},
\batitle{{Visible Emission Line Coronagraph on Aditya-L1}}.
\bjtitle{Current Science}
\bvolume{113},
\bfpage{613}.
\doiurl{https://doi.org/10.18520/cs/v113/i04/613-615}.
\end{barticle}
\endbibitem

\bibitem[\protect\citeauthoryear{{Renotte} et~al.}{2014}]{Proba3}
\begin{bchapter}
\bauthor{\bsnm{{Renotte}}, \binits{E.}},
\bauthor{\bsnm{{Baston}}, \binits{E.C.}},
\bauthor{\bsnm{{Bemporad}}, \binits{A.}},
\bauthor{\bsnm{{Capobianco}}, \binits{G.}},
\bauthor{\bsnm{{Cernica}}, \binits{I.}},
\bauthor{\bsnm{{Darakchiev}}, \binits{R.}},
\bauthor{\bsnm{{Denis}}, \binits{F.}},
\bauthor{\bsnm{{Desselle}}, \binits{R.}},
\bauthor{\bsnm{{De Vos}}, \binits{L.}},
\bauthor{\bsnm{{Fineschi}}, \binits{S.}},
\bauthor{\bsnm{{Focardi}}, \binits{M.}},
\bauthor{\bsnm{{G{\'o}rski}}, \binits{T.}},
\bauthor{\bsnm{{Graczyk}}, \binits{R.}},
\bauthor{\bsnm{{Halain}}, \binits{J.-P.}},
\bauthor{\bsnm{{Hermans}}, \binits{A.}},
\bauthor{\bsnm{{Jackson}}, \binits{C.}},
\bauthor{\bsnm{{Kintziger}}, \binits{C.}},
\bauthor{\bsnm{{Kosiec}}, \binits{J.}},
\bauthor{\bsnm{{Kranitis}}, \binits{N.}},
\bauthor{\bsnm{{Landini}}, \binits{F.}},
\bauthor{\bsnm{{L{\'e}dl}}, \binits{V.}},
\bauthor{\bsnm{{Massone}}, \binits{G.}},
\bauthor{\bsnm{{Mazzoli}}, \binits{A.}},
\bauthor{\bsnm{{Melich}}, \binits{R.}},
\bauthor{\bsnm{{Mollet}}, \binits{D.}},
\bauthor{\bsnm{{Mosdorf}}, \binits{M.}},
\bauthor{\bsnm{{Nicolini}}, \binits{G.}},
\bauthor{\bsnm{{Nicula}}, \binits{B.}},
\bauthor{\bsnm{{Orlea{\'n}ski}}, \binits{P.}},
\bauthor{\bsnm{{Palau}}, \binits{M.-C.}},
\bauthor{\bsnm{{Pancrazzi}}, \binits{M.}},
\bauthor{\bsnm{{Paschalis}}, \binits{A.}},
\bauthor{\bsnm{{Peresty}}, \binits{R.}},
\bauthor{\bsnm{{Plesseria}}, \binits{J.-Y.}},
\bauthor{\bsnm{{Rataj}}, \binits{M.}},
\bauthor{\bsnm{{Romoli}}, \binits{M.}},
\bauthor{\bsnm{{Thizy}}, \binits{C.}},
\bauthor{\bsnm{{Thom{\'e}}}, \binits{M.}},
\bauthor{\bsnm{{Tsinganos}}, \binits{K.}},
\bauthor{\bsnm{{Wodnicki}}, \binits{R.}},
\bauthor{\bsnm{{Walczak}}, \binits{T.}},
\bauthor{\bsnm{{Zhukov}}, \binits{A.}}:
\byear{2014},
\bctitle{{ASPIICS: an externally occulted coronagraph for PROBA-3: Design
  evolution}}.
In: \bbtitle{Space Telescopes and Instrumentation 2014: Optical, Infrared, and
  Millimeter Wave},
\bsertitle{Proceedings of SPIE}
\bseriesno{9143},
\bfpage{91432M}.
\doiurl{https://doi.org/10.1117/12.2056784}.
\adsurl{2014SPIE.9143E..2MR}.
\end{bchapter}
\endbibitem

\bibitem[\protect\citeauthoryear{Seetha and Megala}{2017}]{ADITYA2017}
\begin{barticle}
\bauthor{\bsnm{Seetha}, \binits{S.}},
\bauthor{\bsnm{Megala}, \binits{S.}}:
\byear{2017},
\batitle{{Aditya-L1 mission}}.
\bjtitle{Current Science}
\bvolume{113},
\bfpage{610}.
\doiurl{https://doi.org/10.18520/cs/v113/i04/ 610 - 612}.
\end{barticle}
\endbibitem

\bibitem[\protect\citeauthoryear{{Stenborg} and
  {Cobelli}}{2003}]{Stenborg2003A&A}
\begin{barticle}
\bauthor{\bsnm{{Stenborg}}, \binits{G.}},
\bauthor{\bsnm{{Cobelli}}, \binits{P.J.}}:
\byear{2003},
\batitle{{A wavelet packets equalization technique to reveal the multiple
  spatial-scale nature of coronal structures}}.
\bjtitle{\aap}
\bvolume{398},
\bfpage{1185}.
\doiurl{https://doi.org/10.1051/0004-6361:20021687}.
\adsurl{2003A&A...398.1185S}.
\end{barticle}
\endbibitem

\bibitem[\protect\citeauthoryear{{Stenborg}, {Vourlidas}, and
  {Howard}}{2008}]{Stenborg2008ApJ}
\begin{barticle}
\bauthor{\bsnm{{Stenborg}}, \binits{G.}},
\bauthor{\bsnm{{Vourlidas}}, \binits{A.}},
\bauthor{\bsnm{{Howard}}, \binits{R.A.}}:
\byear{2008},
\batitle{{A Fresh View of the Extreme-Ultraviolet Corona from the Application
  of a New Image-Processing Technique}}.
\bjtitle{\apj}
\bvolume{674},
\bfpage{1201}.
\doiurl{https://doi.org/10.1086/525556}.
\adsurl{2008ApJ...674.1201S}.
\end{barticle}
\endbibitem

\bibitem[\protect\citeauthoryear{{Thernisien}}{2011}]{Thernisien2011ApJS}
\begin{barticle}
\bauthor{\bsnm{{Thernisien}}, \binits{A.}}:
\byear{2011},
\batitle{{Implementation of the Graduated Cylindrical Shell Model for the
  Three-dimensional Reconstruction of Coronal Mass Ejections}}.
\bjtitle{\apjs}
\bvolume{194},
\bfpage{33}.
\doiurl{https://doi.org/10.1088/0067-0049/194/2/33}.
\adsurl{2011ApJS..194...33T}.
\end{barticle}
\endbibitem

\bibitem[\protect\citeauthoryear{{Thernisien}, {Vourlidas}, and
  {Howard}}{2009}]{Thernisien2009SoPh}
\begin{barticle}
\bauthor{\bsnm{{Thernisien}}, \binits{A.}},
\bauthor{\bsnm{{Vourlidas}}, \binits{A.}},
\bauthor{\bsnm{{Howard}}, \binits{R.A.}}:
\byear{2009},
\batitle{{Forward Modeling of Coronal Mass Ejections Using STEREO/SECCHI
  Data}}.
\bjtitle{\solphys}
\bvolume{256},
\bfpage{111}.
\doiurl{https://doi.org/10.1007/s11207-009-9346-5}.
\adsurl{2009SoPh..256..111T}.
\end{barticle}
\endbibitem

\bibitem[\protect\citeauthoryear{{Thernisien}, {Howard}, and
  {Vourlidas}}{2006}]{Thernisien2006ApJ}
\begin{barticle}
\bauthor{\bsnm{{Thernisien}}, \binits{A.F.R.}},
\bauthor{\bsnm{{Howard}}, \binits{R.A.}},
\bauthor{\bsnm{{Vourlidas}}, \binits{A.}}:
\byear{2006},
\batitle{{Modeling of Flux Rope Coronal Mass Ejections}}.
\bjtitle{\apj}
\bvolume{652},
\bfpage{763}.
\doiurl{https://doi.org/10.1086/508254}.
\adsurl{2006ApJ...652..763T}.
\end{barticle}
\endbibitem

\bibitem[\protect\citeauthoryear{{Thompson} et~al.}{2010}]{Thompson2010SoPh}
\begin{barticle}
\bauthor{\bsnm{{Thompson}}, \binits{W.T.}},
\bauthor{\bsnm{{Wei}}, \binits{K.}},
\bauthor{\bsnm{{Burkepile}}, \binits{J.T.}},
\bauthor{\bsnm{{Davila}}, \binits{J.M.}},
\bauthor{\bsnm{{St. Cyr}}, \binits{O.C.}}:
\byear{2010},
\batitle{{Background Subtraction for the SECCHI/COR1 Telescope Aboard STEREO}}.
\bjtitle{\solphys}
\bvolume{262},
\bfpage{213}.
\doiurl{https://doi.org/10.1007/s11207-010-9513-8}.
\adsurl{2010SoPh..262..213T}.
\end{barticle}
\endbibitem

\bibitem[\protect\citeauthoryear{{Woo}}{2005}]{Woo2005SoPh}
\begin{barticle}
\bauthor{\bsnm{{Woo}}, \binits{R.}}:
\byear{2005},
\batitle{{Relating White-Light Coronal Images to Magnetic Fields and Plasma
  Flow}}.
\bjtitle{\solphys}
\bvolume{231},
\bfpage{71}.
\doiurl{https://doi.org/10.1007/s11207-005-1580-x}.
\adsurl{2005SoPh..231...71W}.
\end{barticle}
\endbibitem

\end{thebibliography}

\IfFileExists{\jobname.bbl}{} {\typeout{}
\typeout{****************************************************}
\typeout{****************************************************}
\typeout{** Please run "bibtex \jobname" to obtain} \typeout{**
the bibliography and then re-run LaTeX} \typeout{** twice to fix
the references !}
\typeout{****************************************************}
\typeout{****************************************************}
\typeout{}}

\end{article} 
\end{document}